\documentclass[%
 aip,
 sd,%
 amsmath,
 amssymb,
 preprint,%
]{revtex4-1}

\usepackage{graphicx}
\usepackage{dcolumn}
\usepackage{bm}
\usepackage{subfig}

\usepackage{color}

\begin{document}

\preprint{J. Chem. Phys}

\title [Molecular and colloidal electrostatics] {A robust and accurate formulation of molecular and colloidal electrostatics}

\author{Qiang Sun}
 \email{Qiang.Sun@unimelb.edu.au}
\affiliation{Particulate Fluids Processing Center, Department of Chemical and Biomolecular Engineering, The University of Melbourne, Parkville, VIC 3010, Australia}

\author{Evert Klaseboer}
 \email{evert@ihpc.a-star.edu.sg}
\affiliation{Institute of High Performance Computing, 1 Fusionopolis Way, Singapore 138632, Singapore}

\author{Derek Y. C. Chan}
\email{D.Chan@unimelb.edu.au}
\affiliation{Particulate Fluids Processing Center, School of Mathematics and Statistics, The University of Melbourne, Parkville, VIC 3010, Australia\, and \\Department of Chemistry and Biochemistry, Swinburne University of Technology, Hawthorn VIC 3122, Australia}

\date{\today}

\begin{abstract}
This paper presents a re-formulation of the boundary integral method for the Debye-H$\ddot{\text{u}}$ckel model of molecular and colloidal electrostatics that removes the mathematical singularities that have to date been accepted as an intrinsic part of the conventional boundary integral equation method. The essence of the present boundary regularized integral equation formulation (BRIEF) consists of subtracting a known solution from the conventional boundary integral method in such a way as to cancel out the singularities associated with the Green's function. This approach better reflects the non-singular physical behavior of the systems on boundaries with the benefits of (i) the surface integrals can be evaluated accurately using quadrature without any need to devise special numerical integration procedures, (ii) being able to use quadratic or spline function surface elements to represent the surface more accurately and the variation of the functions within each element is represented to a consistent level of precision by appropriate interpolation functions, (iii) being able to calculate electric fields, even at boundaries, accurately and directly from the potential without having to solve hypersingular integral equations and this imparts high precision in calculating the Maxwell stress tensor and consequently, intermolecular or colloidal forces, (iv) a reliable way to handle geometric configurations in which different parts of the boundary can be very close together without being affected by numerical instabilities, therefore potentials, fields and forces between surfaces can be found accurately at surface separations down to near contact, and (v) having the simplicity of a formulation that does not require complex algorithms to handle singularities will result in significant savings in coding effort and in the reduction of opportunities for coding errors. These advantages are illustrated using examples drawn from molecular and colloidal electrostatics.
\end{abstract}

\maketitle

%
%
%
%
%

%
%
%

%
%
\section{\label{sec:intro}Introduction}

Quantitative estimates of the electrostatic interaction between constituents in molecular and colloidal systems are central to understanding the role of structures and functions in areas ranging from biology, engineering to material science. Specific applications include understanding the translocation of DNA molecules in external fields \cite{doyle2006} or in the vicinity of membranes \cite{buyukdagli2014}, estimating the energetics of protein complexation \cite{kim2010}, modeling biomembrane mechanics \cite{miloshevsky2010}, quantifying the effects of induced charges at dielectric boundaries in soft condensed matter \cite{tyagi2010}, modeling charge transfer energies in electric cells \cite{xiao2014}, electrolyte theory \cite{lange2011} and simulation studies of colloidal systems \cite{alvarez2010}. In the majority of cases, the focus is on the forces and energetics of systems comprised of charged components in an aqueous electrolyte. For general qualitative understanding and in many instances even sufficient for quantitative precision, the linear Debye-H$\ddot{\text{u}}$ckel or the linearised Poisson-Boltzmann model has served as an informative and tractable starting point that can capture most of the important physical ingredients. This has been recognized as such since the classic contributions of Kirkwood \cite{kirkwood1934} on the solvation energies of ionic species and of the Derjaguin-Landau-Verwey-Overbeek (DLVO) model \cite{derjaguin1941, verwey1948} of the interaction between colloidal particles developed in the early part of the last century. Even at present, the Kirkwood model of charged molecules treated as point charges embedded in a continuum dielectric body that is immersed in an implicit continuum solvent model electrolyte is still very much in use for understanding the interaction between complex biological molecules. Furthermore, the available analytic solutions for simple geometries are valuable benchmarks for complex computational methods in molecular electrostatics. Similarly the paradigm introduced by the DLVO theory still underpins current understanding in colloidal interactions.

Understandably, the Kirkwood and DLVO models are based on simple geometries such as spheres, cylinders or planes for which analytical solutions are available for the Debye-H$\ddot{\text{u}}$ckel equation for the electrostatic potential $\phi(\boldsymbol{x})$ in an electrolyte:
\begin{align} \label{eq:DH}
  \nabla^2 \phi(\boldsymbol{x}) - \kappa^{2} \phi(\boldsymbol{x}) = 0
\end{align}
where the ionic concentration of the electrolyte is characterized by the Debye length, $1/\kappa$. For general geometries, the partial differential equation, Eq.~(\ref{eq:DH}), has to be solved numerically using finite difference methods, finite element methods or boundary element methods. 

The finite difference methods and finite element methods are based on discretization of the 3D domain. This approach requires appropriate choice of variable grid resolution that can faithfully represent complex surface shapes and surface spacings as well as handling infinite domains. For complex geometries, this can be challenging. However, these methods generate large but sparse matrix equations for which well-tested computational algorithms are available to handle the numerical task.

In contrast, the boundary element method uses Green's identity to express the solution of Eq.~(\ref{eq:DH}) in terms of $\phi(\boldsymbol{x})$ and its normal derivative $\partial \phi(\boldsymbol{x})/\partial n \equiv \boldsymbol{n}(\boldsymbol{x}) \cdot \nabla \phi(\boldsymbol{x})$ on the 2D boundary, $S$, of the 3D domain where $\boldsymbol{n}(\boldsymbol{x})$ is the outward unit normal of $S$ at $\boldsymbol{x}$. The values of these functions on the boundary are obtained by solving the surface integral equation \cite{becker1992}
\begin{align}\label{eq:CBIM}
c_{0}\phi(\boldsymbol{x}_{0}) + \int_{S} \phi(\boldsymbol{x}) \frac{\partial{G(\boldsymbol{x},\boldsymbol{x}_{0})}}{\partial{n}}\text{ d}S(\boldsymbol{x}) = \int_{S} G(\boldsymbol{x},\boldsymbol{x}_{0}) \frac{\partial{\phi(\boldsymbol{x})}}{\partial{n}}\text{ d}S(\boldsymbol{x})
\end{align}
where the points $\boldsymbol{x}$ and $\boldsymbol{x}_0$ both lie on the surface $S$ that may be the surface of a molecule or a colloidal particle. Here 
\begin{align}\label{eq:GREEN}
G(\boldsymbol{x},\boldsymbol{x}_0) = \frac{\exp(-\kappa |\boldsymbol{x} - \boldsymbol{x}_0|)}{|\boldsymbol{x} - \boldsymbol{x}_0|}
\end{align}
is the Green's function:$\nabla^2 G - \kappa^{2} G = - 4 \pi \delta(\boldsymbol{x} - \boldsymbol{x}_0)$ of Eq.~(\ref{eq:DH}). The constant, $c_0$, in Eq.~(\ref{eq:CBIM}) is the solid angle at $\boldsymbol{x}_0$. It is equal to $2 \pi$ if the tangent plane of $S$ at $\boldsymbol{x}_{0}$ is defined, otherwise it has to be calculated from the local geometry \cite{mantic93, sun2015}. As we shall see later, our formulation of the boundary integral equation is independent of the value of $c_0$. Therefore, once the values of $\phi(\boldsymbol{x})$ and $\partial \phi(\boldsymbol{x})/\partial n$ on the surface, $S$, have been found, the value of $\phi(\boldsymbol{x})$ anywhere in the 3D domain can be determined by an integral over the surface \cite{becker1992}.

For an electrostatic problem, either $\phi(\boldsymbol{x})$ or $\partial \phi(\boldsymbol{x})/\partial n$ or a relation between them \cite{carnie1993} on the surface is known from the boundary conditions appropriate to the problem. Thus the unknown quantity in Eq.~(\ref{eq:CBIM}) can be found by solving a problem in 2D. In contrast, finite difference or finite element methods require solving a problem in 3D. This reduction in dimension and hence the number of unknowns, together with the accurate account of conditions at infinity is a desirable tradeoff for a slightly more complex formulation that favors the boundary element approach.

Traditionally, the boundary element method is regarded to have two major disadvantages. Firstly, discretization of Eq.~(\ref{eq:CBIM}) results in a dense matrix equation that has $O(N^3)$ complexity, rendering it unsuited for large problems. However, with the recent advent of fast solvers based on Krylov subspace iteration methods together with Fast Fourier Transform and Singular Value Decomposition algorithms \cite{altman2009} the computational cost can be reduced to a practically competitive level of $O(N \log N)$. Secondly, the Green's function formulation of Eq.~(\ref{eq:CBIM}) contains singularities as $\boldsymbol{x} \rightarrow \boldsymbol{x}_0$ in the two surface integrals that are inherent in the Green's function, $G$, and its normal derivative, $\partial G/\partial n$. Such singular behavior is a consequence of the mathematical formulation and does not relate to any physical divergences on the boundary surfaces. Although these integrals are bounded in spite of the singularities in $G$ and $\partial G/\partial n$, that is, these singularities are integrable, nonetheless,  the mathematical complexity needed to handle such singular behavior means that usually only the simplest area elements are used to discretize $S$. By far the most common approach is to represent $S$ by a mesh of planar triangular area elements and assume that $\phi$ and $\partial \phi/\partial n$ are constants within each element as the unknowns to be determined.

A more serious consequence of the singular nature of $G$ and $\partial G/\partial n$ is that when problems involve having two parts of the boundary being very close together, for example, when the surfaces of two ions or colloidal particles are nearly in contact, such singular behavior will limit the precision for which the surface integrals can be evaluated numerically because the near singular behavior at one surface can adversely affect the precision of the evaluation of integrals on a nearby surface.  

A further problematic consequence of such singular behavior is that in calculating the force between charged entities, it is necessary to determine the electric field, $\boldsymbol{E} = - \nabla \phi$, that enters into the Maxwell stress tensor. It has been suggested that a boundary integral equation for $\boldsymbol{E}$ can be found by taking the gradient of Eq.~(\ref{eq:CBIM}) \cite{lu2005}, resulting in hypersingular integral equations in which the more strongly divergent integrals need to be interpreted as principal value integrals. Numerical evaluation of such integrals requires special care that impacts adversely on the achievable precision. This hypersingular behavior arises from the interchange of integration and differentiation -- an ill-advised procedure for integrals that do not converge absolutely.

Given the achieved advances in the development of $O(N \log N)$ fast solvers for dense linear systems that arise from the boundary integral solution of the Debye-H$\ddot{\text{u}}$ckel equation, it is timely to re-examine the foundation of the integral equation, Eq.~(\ref{eq:CBIM}), and seek to eliminate the mathematical singularities that originate from the Green's function $G(\boldsymbol{x},\boldsymbol{x}_0)$. Such singularities, while long accepted in boundary integral equations as unavoidable, do not have any actual physical basis. Thus success in eliminating them will obviously be very beneficial to many areas in chemical physics in which molecular and colloidal electrostatics feature as key components of a larger framework.

In Secs.~\ref{sec:motivate} and~\ref{sec:BRIEF}, we present a new formulation of the boundary integral equation solution of the Debye-H$\ddot{\text{u}}$ckel equation that does not contain the traditional singularities. The consequences of this non-singular formulation are: 
\begin {enumerate}
\item the surface integrals can be evaluated accurately using quadrature without any need to devise special numerical integration procedures so that, for example, standard Gauss quadrature can be used, 
\item we can use quadratic surface elements to represent the surface $S$ more accurately and the variation of the functions within each element is represented to a consistent level of precision by quadratic interpolation functions; and opens the possibility to use spline functions or higher order functions to represent surfaces,
\item  electric fields, even at boundaries, can be evaluated accurately and directly from the potential without having to solve hypersingular integral equations and this imparts high precision in calculating the Maxwell stress tensor and consequently, intermolecular or colloidal forces, 
\item geometric configurations in which different parts of the boundary are very close together will not cause numerical instabilities, thus potentials, fields and forces between surfaces can be found accurately at surface separations down to near contact,
\item the simplicity of the formulation in not requiring complex algorithms to handle singularities means significant savings in coding effort and reduction of opportunities for coding errors, and
\item multiple domains connected by boundary conditions can be implemented with relative ease.
\end {enumerate}

Since the implementation of fast $O(N \log N)$ algorithms is well-developed and documented, we will only focus on the non-singular formulation of relevant physical problems that will be called the boundary regularized integral equation formulation - BRIEF, and we present numerical examples that will highlight the precision that our approach can furnish. 

\section{\label{sec:motivate}Motivation}
To provide background to our boundary regularized integral equation formulation (BRIEF) for the Debye-H$\ddot{\text{u}}$ckel equation, we first consider the boundary integral formulation of the solution of the Laplace equation for $\phi$:
\begin{align} \label{eq:LaplaceDE_phi}
  \nabla^2 \phi(\boldsymbol{x}) = 0
\end{align}
for which the standard boundary integral equation on the surface $S$ that encloses the solution domain is~\cite{becker1992}
\begin{align}\label{eq:LaplaceCBIM_phi}
c_{0}\phi(\boldsymbol{x}_{0}) + \int_{S} \phi(\boldsymbol{x}) \frac{\partial{G_0(\boldsymbol{x},\boldsymbol{x}_{0})}}{\partial{n}}\text{ d}S(\boldsymbol{x}) = \int_{S} G_0(\boldsymbol{x},\boldsymbol{x}_{0}) \frac{\partial{\phi(\boldsymbol{x})}}{\partial{n}}\text{ d}S(\boldsymbol{x})
\end{align}
where the Green's function for the Laplace equation (\ref{eq:LaplaceDE_phi}) is
\begin{align}\label{eq:LaplaceGREEN}
G_0(\boldsymbol{x},\boldsymbol{x}_0) = \frac{1}{|\boldsymbol{x} - \boldsymbol{x}_0|}.
\end{align}

The standard way to ameliorate the singularity in ${\partial G_0}/{\partial n}$ on the LHS of Eq. (\ref{eq:LaplaceCBIM_phi}) is to note that the function $[\phi(\boldsymbol{x}) - \phi(\boldsymbol{x}_0)]$ also satisfies the Laplace equation, Eq. (\ref{eq:LaplaceDE_phi}), and therefore the corresponding standard boundary integral equation for $[\phi(\boldsymbol{x}) - \phi(\boldsymbol{x}_0)]$ is
\begin{align}\label{eq:LaplaceCBIM_phi_subtracted}
\int_{S} [\phi(\boldsymbol{x}) - \phi(\boldsymbol{x}_0)] \frac{\partial{G_0(\boldsymbol{x},\boldsymbol{x}_{0})}}{\partial{n}}\text{ d}S(\boldsymbol{x}) = \int_{S} G_0(\boldsymbol{x},\boldsymbol{x}_{0}) \frac{\partial{\phi(\boldsymbol{x})}}{\partial{n}}\text{ d}S(\boldsymbol{x}).
\end{align}
Thus provided the function $\phi(\boldsymbol{x})$ is continuous at $\boldsymbol{x}_0$, the use of Eq. (\ref{eq:LaplaceCBIM_phi_subtracted}) instead of Eq. (\ref{eq:LaplaceCBIM_phi}) means that the remaining integrable weak singularities can be accommodated by a local change of variables when $\boldsymbol{x} \rightarrow \boldsymbol{x}_0$.

We now extend this approach to the case when $\phi(\boldsymbol{x})$ satisfies the Debye-H$\ddot{\text{u}}$ckel equation.

\section{\label{sec:BRIEF}Formulation}
For ease of future reference, we designate the boundary integral equation in Eq.~(\ref{eq:CBIM}) derived from the conventional boundary integral method as CBIM. Our objective is to remove analytically the singularities in $G$ and $\partial G/\partial n$ in Eq.~(\ref{eq:CBIM}), that occur as $\boldsymbol {x} \rightarrow \boldsymbol{x}_{0}$.  Such singularities are the consequence of the mathematical derivation of  Eq.~(\ref{eq:CBIM}) and have no intrinsic physical basis. It is also worth noticing that the singular behavior of $G$ is in fact identical to $G_{0} \equiv 1/|\boldsymbol {x} - \boldsymbol{x}_{0}|$ since $G \equiv G_{0} + \Delta G $ where $ \Delta G \equiv  [\exp (-\kappa |\boldsymbol{x} - \boldsymbol{x}_0| ) - 1]/|\boldsymbol{x} - \boldsymbol{x}_0| $ is finite as $\boldsymbol {x} \rightarrow \boldsymbol{x}_{0}$. The same analysis can be applied to $\partial G/\partial n$ as well. The approach that we use to deal with these singularities, a simple version of which is given in Sec.~\ref{sec:motivate}, is adapted from de-singularisation of integral equations that occur in fluid mechanics, elasticity and acoustics \cite{klaseboer2012, sun2014, sun2015}. 

We begin by first considering the function $\psi(\boldsymbol{x})$ that satisfies the Laplace equation for the same domain as where Eq. (\ref{eq:DH}) is valid:
\begin{align} \label{eq:LaplaceDE}
  \nabla^2 \psi(\boldsymbol{x}) = 0.
\end{align}
The corresponding conventional boundary integral equation for $\psi(\boldsymbol{x})$ for the same surface $S$ is
\begin{align}\label{eq:LaplaceCBIM}
c_{0}\psi(\boldsymbol{x}_{0}) + \int_{S} \psi(\boldsymbol{x}) \frac{\partial{G_0(\boldsymbol{x},\boldsymbol{x}_{0})}}{\partial{n}}\text{ d}S(\boldsymbol{x}) = \int_{S} G_0(\boldsymbol{x},\boldsymbol{x}_{0}) \frac{\partial{\psi(\boldsymbol{x})}}{\partial{n}}\text{ d}S(\boldsymbol{x})
\end{align}
where $G_0$ is given by Eq. (\ref{eq:LaplaceGREEN}). In order to use $\psi(\boldsymbol{x})$ to remove the singularities in Eq.~(\ref{eq:CBIM}), we require it to assume the following special values at $\boldsymbol{x} = \boldsymbol{x}_0$:
\begin{align} \label{eq:LaplaceBCphi}
  \psi(\boldsymbol{x}_0) &= \phi(\boldsymbol{x}_0) \\ \nonumber
  \\ 
  \frac{\partial{\psi(\boldsymbol{x}_0)}}{\partial{n}} &= \frac{\partial{\phi(\boldsymbol{x}_0)}}{\partial{n}}. \label{eq:LaplaceBCdphidn}
\end{align}
Thus subtracting Eq.~(\ref{eq:LaplaceCBIM}) from Eq.~(\ref{eq:CBIM}) we can eliminate the term $c_0 \phi(\boldsymbol{x}_0)$ to give
\begin{align}\label{eq:DiffCBIM}
\nonumber
\int_{S}  \left[ \phi(\boldsymbol{x}) \frac{\partial{G(\boldsymbol{x},\boldsymbol{x}_{0})}}{\partial{n}} - \psi(\boldsymbol{x}) \frac{\partial{G_0(\boldsymbol{x},\boldsymbol{x}_{0})}}{\partial{n}}  \right] & \text{ d}S(\boldsymbol{x}) = \\
  \int_{S}  & \left[ G(\boldsymbol{x},\boldsymbol{x}_{0}) \frac{\partial{\phi(\boldsymbol{x})}}{\partial{n}} - G_0(\boldsymbol{x},\boldsymbol{x}_{0}) \frac{\partial{\psi(\boldsymbol{x})}}{\partial{n}} \right] \text{ d}S(\boldsymbol{x}).
\end{align}
To satisfy Eqs.~(\ref{eq:LaplaceBCphi}) and (\ref{eq:LaplaceBCdphidn}), we can choose $\psi(\boldsymbol{x})$ to have the form
\begin{align}\label{eq:psi0fg}
\psi(\boldsymbol{x}) = \phi(\boldsymbol{x}_0) g(\boldsymbol{x}) + \frac{\partial{\phi(\boldsymbol{x}_0)}}{\partial{n}} f(\boldsymbol{x})
\end{align}
where $g(\boldsymbol{x})$ and $f(\boldsymbol{x})$ satisfy the Laplace equation with the following conditions at $\boldsymbol{x} = \boldsymbol{x}_0$
\begin{align}  
\nabla^2 g(\boldsymbol{x}) &= 0,  \quad  g(\boldsymbol{x}_0) = 1,  \quad  
\nabla g(\boldsymbol{x}_0) \cdot \boldsymbol{n}(\boldsymbol{x}_0) = 0,  \label{eq:gEqn}
\\
 \nabla^2 f(\boldsymbol{x}) &= 0,  \quad  f(\boldsymbol{x}_0) = 0,  \quad  
 \nabla f(\boldsymbol{x}_0) \cdot \boldsymbol{n}(\boldsymbol{x}_0) = 1,  \label{eq:fEqn}
\end{align}
and this will ensure both integrands in Eq.~(\ref{eq:DiffCBIM}) will not be singular at $\boldsymbol{x} = \boldsymbol{x}_0$. Note that the conditions on $g(\boldsymbol{x})$ and $f(\boldsymbol{x})$ in Eqs. (\ref{eq:gEqn}) and (\ref{eq:fEqn}) are constraints on at a single position $\boldsymbol{x} = \boldsymbol{x}_0$ and are not general boundary data on a surface.

Finally, we have the key result for our boundary regularized integral equation formulation (BRIEF):
\begin{align}\label{eq:BRIEF}
\nonumber
\int_{S}  \left[  \phi(\boldsymbol{x}) \frac{\partial{G(\boldsymbol{x},\boldsymbol{x}_{0})}}{\partial{n}} - \phi(\boldsymbol{x}_0) g(\boldsymbol{x}) \frac{\partial{G_0(\boldsymbol{x},\boldsymbol{x}_{0})}}{\partial{n}}    + \phi(\boldsymbol{x}_0)  \frac{\partial{g(\boldsymbol{x})}}{\partial{n}}G_{0}(\boldsymbol{x},\boldsymbol{x}_{0}) \right]\text{ d}S(\boldsymbol{x}) = \\ 
 \int_{S} \left[  \frac{\partial{\phi(\boldsymbol{x})}}{\partial{n}} G(\boldsymbol{x},\boldsymbol{x}_{0})  - \frac{\partial{\phi(\boldsymbol{x}_0)}}{\partial{n}}  \frac{\partial{f(\boldsymbol{x})}}{\partial{n}} G_0(\boldsymbol{x},\boldsymbol{x}_{0})  +\frac{\partial{\phi(\boldsymbol{x}_0)}}{\partial{n}} f(\boldsymbol{x}) \frac{\partial{G_0(\boldsymbol{x},\boldsymbol{x}_{0})}}{\partial{n}} \right] \text{ d}S(\boldsymbol{x}). 
\end{align}
It can be shown that this integral equation contains no singularities provided $g(\boldsymbol{x})$ and $f(\boldsymbol{x})$ have simple mathematical smoothness properties,\cite{klaseboer2012} and so Eq.~(\ref{eq:BRIEF}) can be solved by straightforward numerical methods. It replaces Eq.~(\ref{eq:CBIM}), the equation from the conventional boundary integral method (CBIM) that contains singular integrands~\cite{altman2009,lu2005,yoon1990,cooper2014,juffer1991}. This equation connecting $\phi$ and $\partial \phi / \partial n$ on the surface $S$ is the starting point of our boundary regularized integral equation formulation (BRIEF) of molecular and colloidal electrostatics. As we shall see, the absence of mathematical singularities in the BRIEF simplifies numerical implementation and coding effort resulting in very significant improvements in numerical precision even for problems that pose nearly insurmountable difficulties for the CBIM.

There are many possible choices for $g(\boldsymbol{x})$ and $f(\boldsymbol{x})$ satisfying Eqs.~(\ref{eq:gEqn}) and (\ref{eq:fEqn}) that will ensure Eq.~(\ref{eq:BRIEF}) is non-singular. For instance, we can take
\begin{align} \label{eq:simpleFandG}
    g(\boldsymbol{x}) = 1, \qquad  f(\boldsymbol{x}) = \boldsymbol{n}(\boldsymbol{x}_0) \cdot (\boldsymbol{x} - \boldsymbol{x}_0).
\end{align}
Although for this simple choice, $\psi(\boldsymbol{x})$ does not vanish at infinity for external problems, its integral over the closed surface at infinity can be found analytically and its magnitude is equal to $4 \pi \phi(\boldsymbol{x}_0)$~\cite{klaseboer2012, sun2015} while its sign depends on the direction of the normal vector. However, there are other choices that may better suit the problem at hand \cite{sun2014}.

Although we have presented the derivation of the BRIEF by considering an integral equation on a single surface, $S$, the generalisation to more complex surface topologies and multiple domains is straightforward \cite{altman2009}, see also the examples below.

\section{Implementation and Examples}

To illustrate the utility and precision that can be achieved by our boundary regularized integral equation formulation (BRIEF) we benchmark against results from recent approaches using the conventional boundary integral method (CBIM) to calculate the solvation energy of the venerable Kirkwood model of an ion in electrolyte. For applications in colloidal electrostatics, we compare against analytic results for the interaction between two spheres available as infinite series expansions in terms of orthogonal functions. We also present illustrative examples of systems with surfaces that are almost in contact for which the geometry can be quite challenging for the CBIM because of the presence of singularities in the integral equation but in contrast, pose no numerical problems using BRIEF because of the absence of singularities in the integrands.
 
\subsection{Molecular electrostatics}

\subsubsection{Kirkwood ion}

\begin{figure} 
  \centering
  \subfloat[]{ \includegraphics[width=0.45\textwidth]{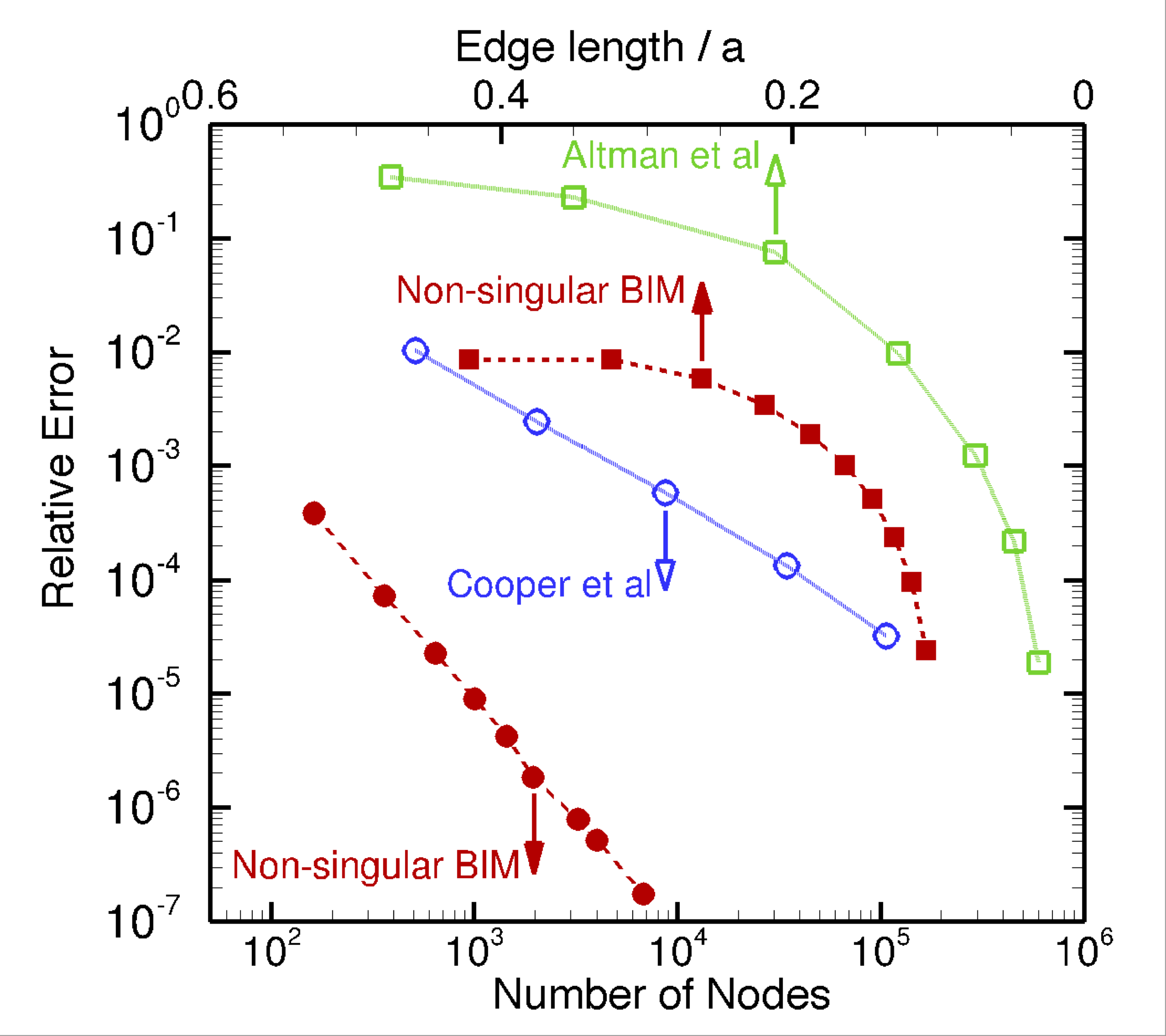} }
  \subfloat[]{ \includegraphics[width=0.45\textwidth]{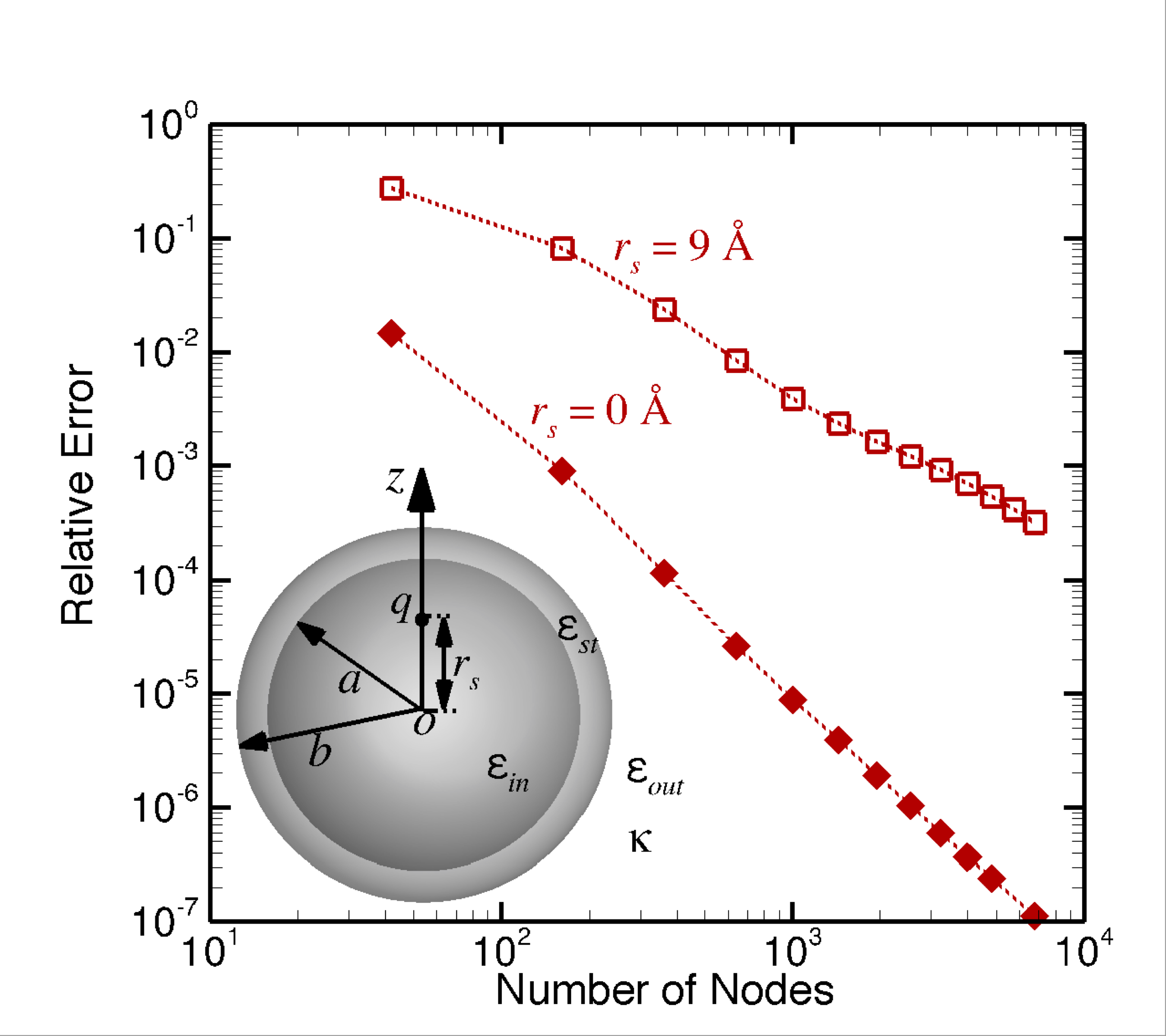} } 
\caption{\label{fig:Kirkwood}  (Color) Variation of the relative error in the solvation energy, $E_{solv}$ with the number of nodes (upper horizontal axis and up arrows) or the edge length (lower horizontal axis and down arrows) of the surface area elements for a Kirkwood ion comprised of a point charge embedded at position $r_s$ inside a sphere of radius, $a$, dielectric constant, $\epsilon_{in} = 4$ immersed in a solvent of dielectric constant, $\epsilon_{out} = 80$ using our present method BRIEF compared to results of (a) Cooper \emph{et al.} \cite{cooper2014} with $a = 4$ \AA, $r_s = 0$, in a dielectric solvent without a Stern layer ($\kappa = 0$ and $b = a$) and of Altman \emph{et el.} \cite{altman2009} with $a = 20$ \AA, $b = 22$ \AA, $\epsilon_{st} = 80$, $r_s = 18$ \AA, in an electrolyte $1/\kappa = 8$ \AA; and (b) for $a = 10$ \AA, $b = 12$ \AA, $\epsilon_{st} = 8$, $1/\kappa = 8$ \AA, showing how the relative error of the solvation energy varies with the position, $r_s$ of the embedded charge. The reference analytic results for the solvation energy in these examples are given in Appendix \ref{appx:kirkwoodsol}.}
\end{figure}

To benchmark against known analytic results we consider calculations of the solvation energy of a Kirkwood ion \cite{kirkwood1934} that was the first molecular electrostatics model on which the conventional boundary integral method was tested\cite{yoon1990}. The model comprises a point charge, $q$ embedded at a location $\boldsymbol{x}_s = (0,0,r_s)$ from the center of a sphere of radius, $a~(>r_s)$ and dielectric constant, $\epsilon_{in}$, immersed in a solvent of dielectric constant, $\epsilon_{out}$. The solvent can also be an electrolyte characterized by a Debye length, $1/\kappa$. In addition, the spherical ion can have a concentric outer shell or Stern layer, of radius, $b$, that excludes ionic species in the thin layer, $a < r < b$. The Stern layer may also have a different dielectric constant, $\epsilon_{st}$. A schematic representation of the Kirkwood ion is given in Fig.~\ref{fig:Kirkwood}b.

The potential $\phi$ is determined by the following equations in different spatial domains
\begin{eqnarray}  
 \nabla^2 \phi(\boldsymbol{x}) 
  &=&    - (q/\epsilon_0 \epsilon_{in}) \; \delta(\boldsymbol{x} - \boldsymbol{x}_s),  \quad 0 < r < a \label{eq:phi_de_a} \\
  &=&  \qquad  \qquad 0,  \qquad \qquad \qquad a < r < b  \label{eq:phi_de_b} \\
  &=&  \qquad  \kappa^2 \phi(\boldsymbol{x}) , \qquad \qquad \qquad  r > b.   \label{eq:phi_de_c}
  \end{eqnarray}
The solution inside the ion $0 < r < a$ can be written as the sum of the coulomb potential due to the point ion and a reaction potential
\begin{align} \label{eq:phi_react}
 \phi(\boldsymbol{x}) =\frac{q}{4 \pi \epsilon_0 \epsilon_{in}} \frac{1}{|\boldsymbol{x} - \boldsymbol{x}_s|}  \; +\;  \phi_{react}(\boldsymbol{x}) , \qquad 0 < r < a.
\end{align}
At the boundaries $r = a$ and $r = b$, we have the continuity conditions of $\phi$ and $\epsilon~\partial \phi / \partial r$, whereas $\phi$ vanishes as $r \rightarrow \infty$. The solution expressed in terms of infinite series is given in Appendix \ref{appx:kirkwoodsol}.

The solvation energy, $E_{solv}$ is given in terms of the reaction potential evaluated at the location of the point ion at $\boldsymbol{x} = \boldsymbol{x}_s$ \cite{kirkwood1934}
\begin{align} \label{eq:E_solv}
  E_{solv} = \frac{1}{2} q \; \phi_{react}(\boldsymbol{x}_s).
\end{align}

Details of using BRIEF to calculate $\phi_{react}(\boldsymbol{x}_s)$ are given in Appendix B. The surface of the Kirkwood sphere is represented by triangular shaped quadratic elements. The unknowns are potential values at the nodes of the elements. Variation of the potential within each quadratic element is obtained by quadratic interpolation of the nodal values. The surface integral over each element is calculated by quadrature and as the resulting linear system is not large, the linear equations are solved by Gauss elimination.

 In Fig.~\ref{fig:Kirkwood}a we compare the relative error in $E_{solv}$ obtained by BRIEF for a Kirkwood ion to corresponding results obtained by CBIM in a dielectric solvent \cite{cooper2014} and an electrolyte solvent \cite{altman2009} for different mesh size or number of nodes used in the evaluation of the surface integrals. It is evident that the relative error obtained using the BRIEF that has no singular integrals can be 2 orders of magnitude smaller than that of the CBIM, or conversely for the CBIM to achieve the same precision as the BRIEF, over a 100-fold increase in the number of nodes will be required.  These results clearly demonstrate the superior efficiency of the BRIEF.  In Fig.~\ref{fig:Kirkwood}b, we  quantify how the position of the point ion inside the solute sphere can affect the relative error in the solvation energy. We see that even when the ion is located at 1~\AA~from the surface of a 10~\AA~radius solvent sphere, accurate results can still be obtained by the BRIEF with a modest number of nodes. When the distance of the ion from the surface of the ion is comparable to the thickness of the Stern layer, the rate of convergence with respect to the number of nodes is slower.

\subsubsection{Dumbbell zwitterion}

\begin{figure}  
 \centering
 \subfloat[]{ \includegraphics[width=0.45\textwidth]{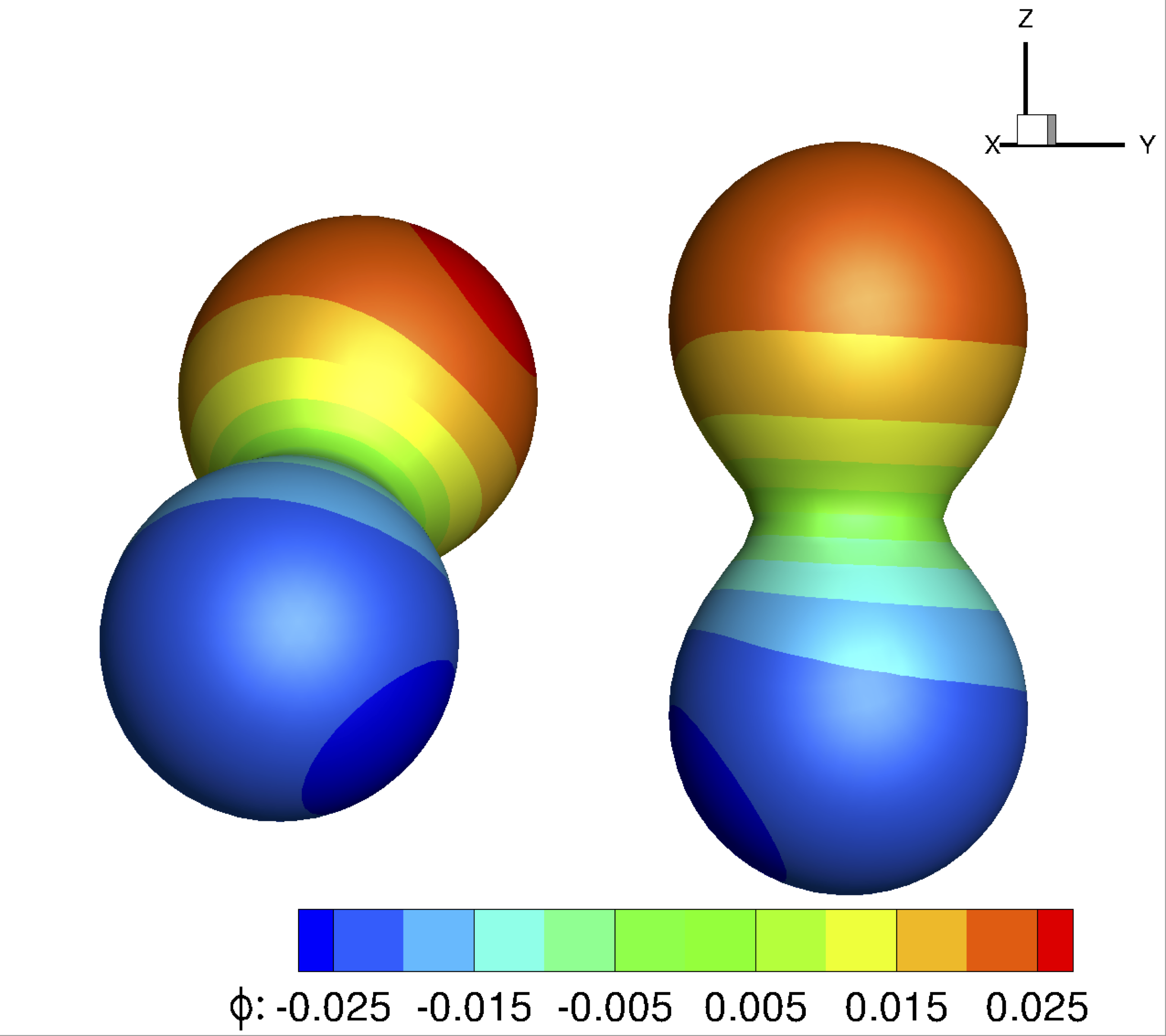} }
 \subfloat[]{ \includegraphics[width=0.45\textwidth]{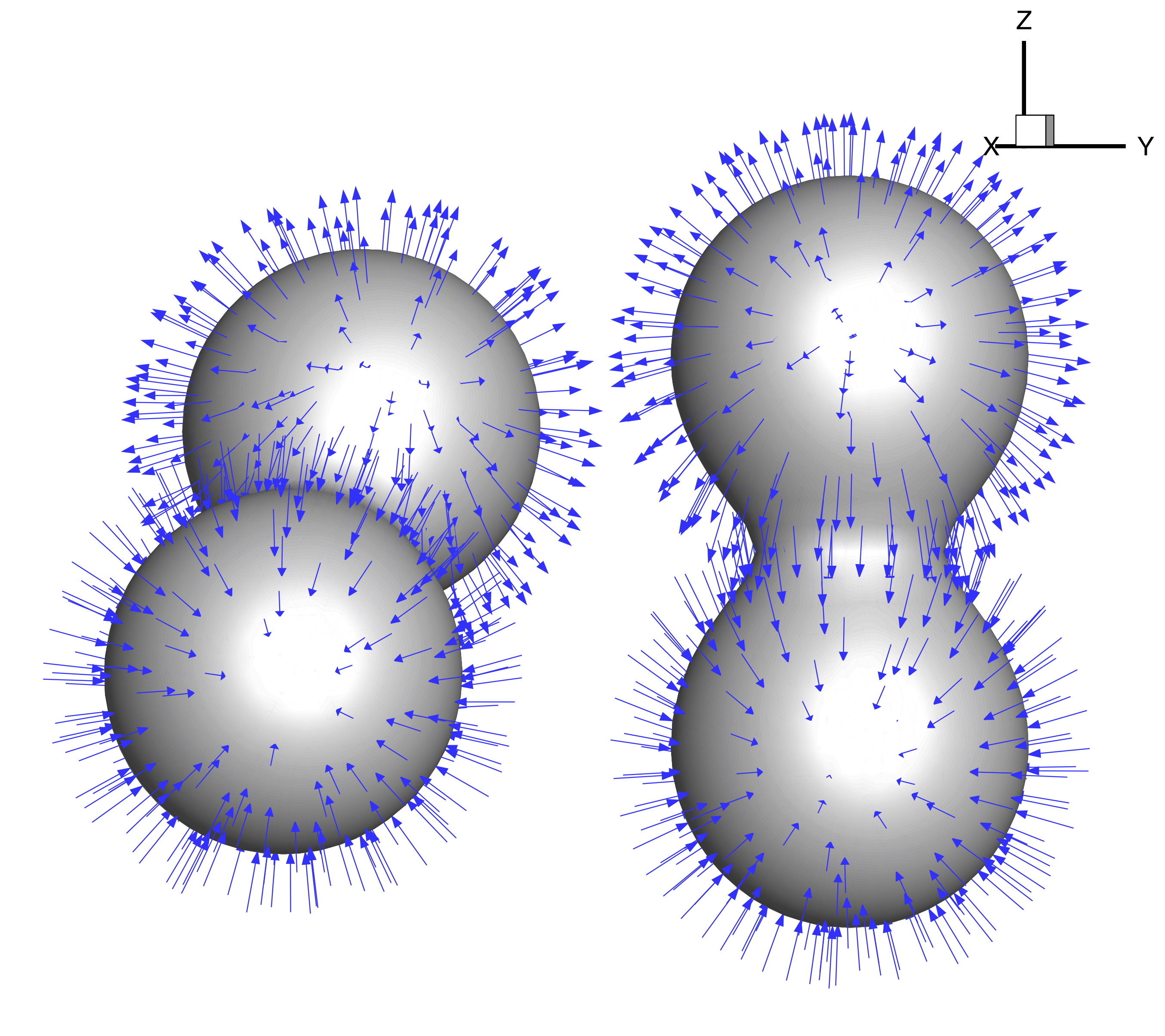} }
 \caption{\label{fig:zwitterion}  (Color) The surface potential $\phi$ in volts (multimedia view) and the electric field at the surface of two dumbbell shapes zwitterionic molecules each of which is comprised a positive and a negative elementary point charge at the foci of the dumbbells of dielectric constant $\epsilon_{in} = 4$ in a solvent of dielectric constant $\epsilon_{out} = 80$. The dumbbells have length $2a = 20$ \AA~and the waist has width $2d = 5$ \AA,  so that $b/a = 0.4757$, $c/a = 0.5303$ and the exterior electrolyte is characterized by $\kappa a = 1.25$. The distance between the axes of the dumbbells is $14.5$ \AA, corresponding to a distance of closest approach of 9.514 \AA. The surface of each dumbbell is represented by a mesh of 1962 nodes connecting 980 quadratic elements.}
\end{figure}

To illustrate how our boundary regularized integral equation formulation (BRIEF) can be used to calculate accurately the potential and electric field $\boldsymbol{E}= -\nabla \phi$ at the surface of charged molecules without using the hyper-singular boundary integral formulations (see Appendix \ref{appx:Efield} for details), we consider an axisymmetric dumbbell shaped zwitterion described in the body axis frame $(X, Y, Z)$ where $Z$ is the axis of rotation and $R^2 = X^2 + Y^2$ by\cite{chwang1974}, (see Fig.~\ref{fig:zwitterion}) 
\begin{align} \label{eq:dumbbell}
  \left[ (Z+c)^{2} + R^2 \right]^{-\frac{3}{2}} + \left[ (Z-c)^2 + R^2 \right]^{-\frac{3}{2}} = 2 \left(c^2+d^2 \right)^{-\frac{3}{2}}.
\end{align}
A point charge, $q$ or $-q$ is placed at each of the foci at $Z=\pm c$ inside the dielectric dumbbell that has dielectric constant $\epsilon_{in}$. The zwitterion is immersed in a continuum electrolyte characterized by dielectric constant $\epsilon_{out}$ and Debye length $1/\kappa$. The dumbbell has length, $2a$, with a narrow neck of width $2d$ at $Z = 0$ in between two lobes of width $2b$.

In Fig.~\ref{fig:zwitterion} we show a pair of zwitterions at separation 9.514 {\AA} between their points of closest approach with  the surface potential $\phi$ on each zwitterion indicated by a color scale. The corresponding external electric vector field on the surface is calculated according to Eq.~(\ref{eq:electricfield}). Also, corresponding to Fig.~\ref{fig:zwitterion}a, we show a video of the variation of the surface potential as the relative orientation of the zwitterions changes (multimedia view).

\subsection{Colloidal electrostatics}

\subsubsection{Potential and field}

We now illustrate the utility of the boundary regularized integral equation formulation (BRIEF) in colloidal electrostatics. In the Debye-H$\ddot{\text{u}}$ckel model, the colloidal particles of dielectric constant, $\epsilon_{in}$, are assumed to carry a specified uniform surface charge density, $\sigma_{ch}$. They interact across an electrolyte characterized by $\epsilon_{out}$ and $\kappa$. Inside the dielectric particle, the electrostatic potential, $\phi$, obeys the Laplace equation and in the outer electrotrolyte, $\phi$ is governed by Eq.~(\ref{eq:DH}). The usual electrostatic boundary conditions at the surface are the continuity of $\phi$ and the condition $[\epsilon~\partial \phi / \partial n]_{in} - [\epsilon~\partial \phi / \partial n]_{out} = \sigma_{ch}/\epsilon_0$. For simple geometries such as for the interaction between two colloidal spheres, it is possible to obtain explicit forms for $\phi$ in terms of infinite series expansions of orthogonal functions \cite{glendinning1983}. The boundary integral method has been used as the starting point of a perturbation calculation \cite{mccartney1969}.

\begin{figure}  
  \centering
  \subfloat[]{ \includegraphics[width=0.45\textwidth]{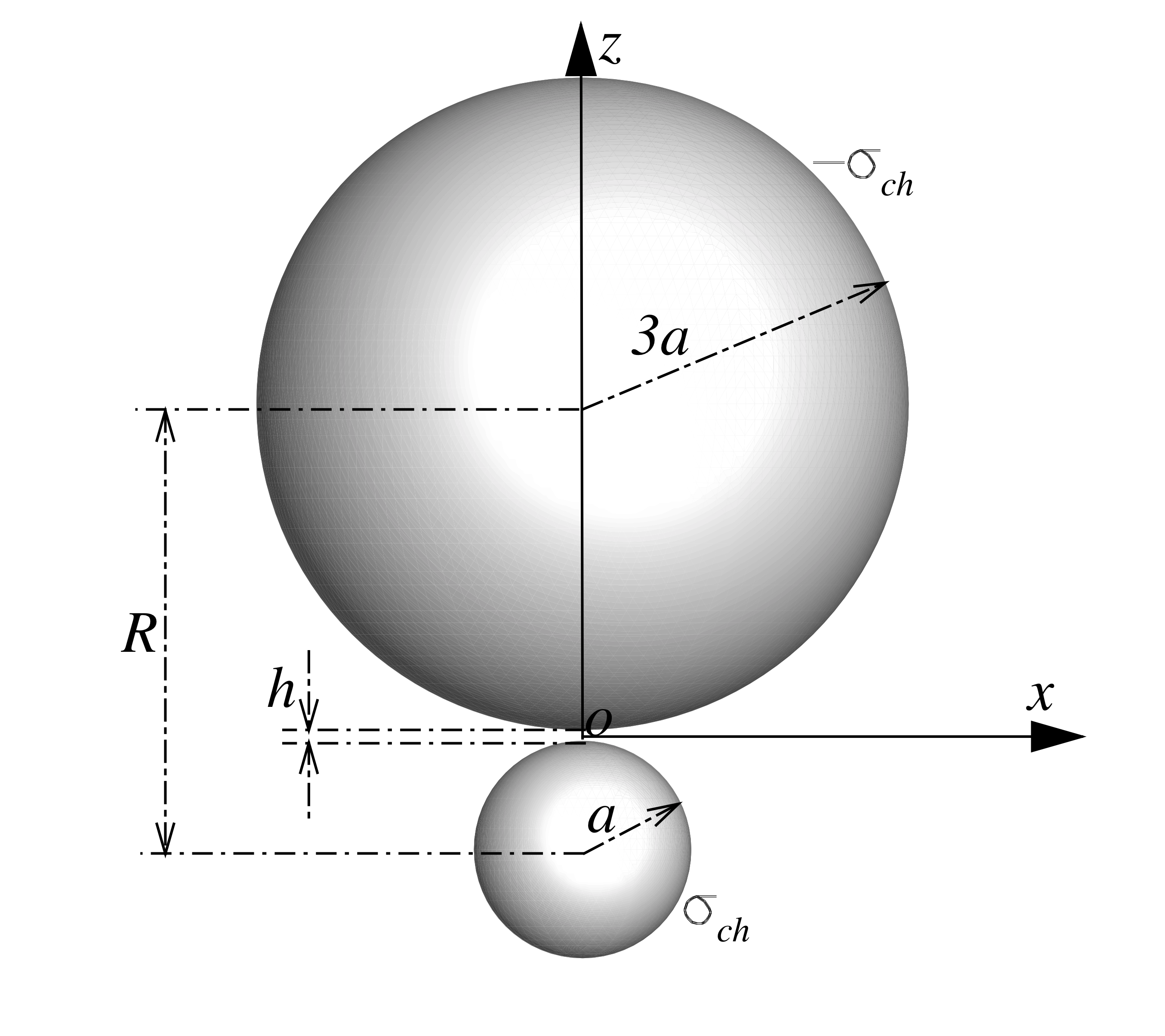} }
  \subfloat[]{ \includegraphics[width=0.45\textwidth]{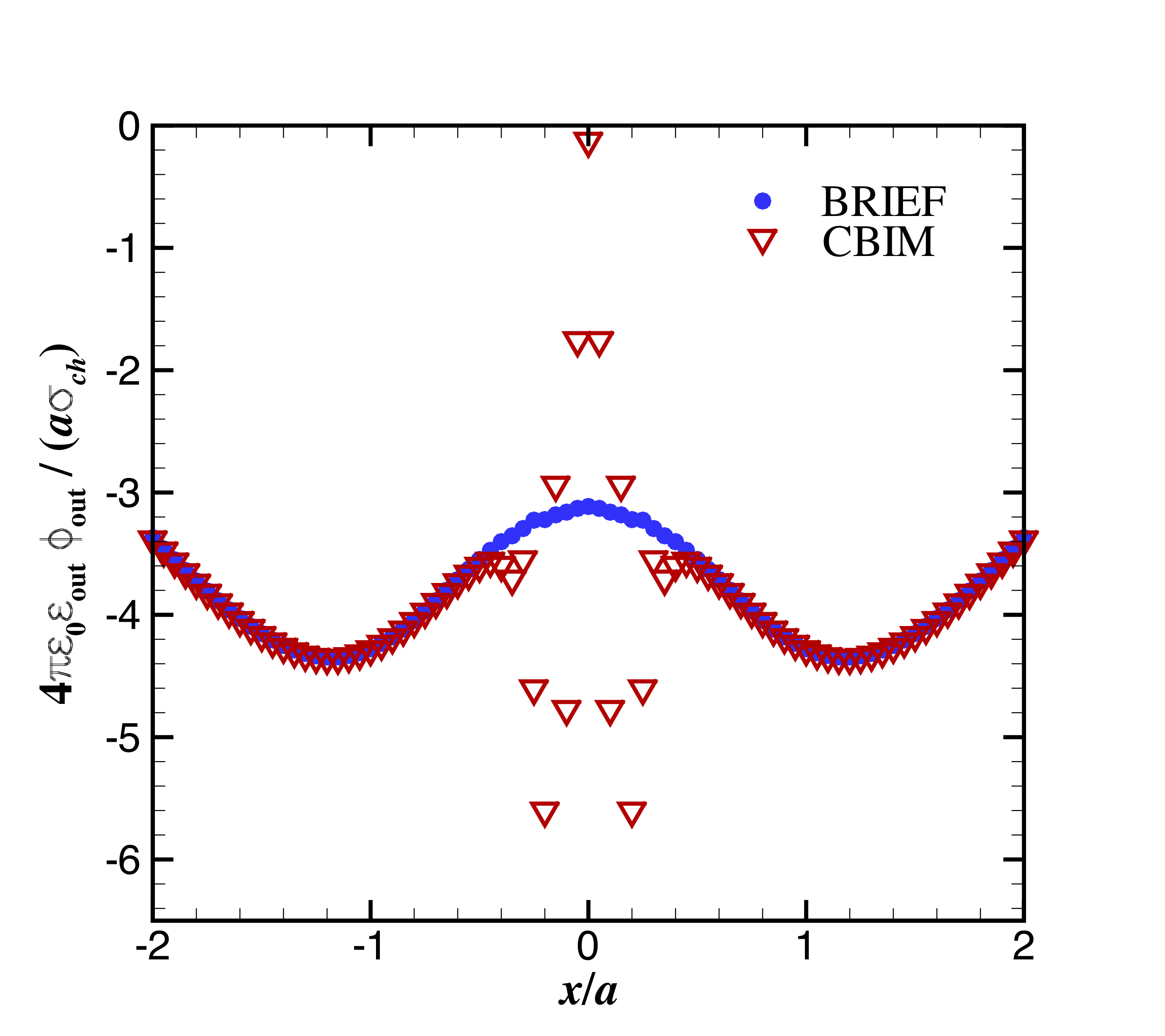} }
\caption{\label{fig:bigsmallspheres} (Color) (a) The configuration of 2 nearly touching spheres at a separation, $h$. The spheres carry equal and opposite surface charge densities with $h/a = 10^{-3}$ and $\kappa a = 1$. (b) Variation of the potential within the median plane $z = 0$ according to the CBIM and the BRIEF. The surface of the smaller sphere is represented by 3072 linear elements comprising 1538 nodes and that of the surface of the larger sphere by 12,288 linear elements comprising 6146 nodes.}
\end{figure}

We have verified that our non-singular boundary regularized integral equation formulation (BRIEF) can reproduce the infinite series solution for the force between two spheres of the same size \cite{glendinning1983}. Here we highlight the advantages of the BRIEF in being able to calculate the potential very accurately in the region between two very nearly touching dielectric spheres - a problem that is very challenging using the series expansion method or the conventional formulation of the boundary integral method. The dielectric spheres with $\epsilon_{in} = 2$, radii $a$ and $3a$ are positioned at a minimum separation $h = a/1000$. The spheres carry equal and opposite uniform surface charge densities, $\pm \sigma_{ch}$ and are immersed in an electrolyte characterized by $\epsilon_{out} = 80$ and $\kappa a = 1$. The centers of the spheres are located along the $z$-axis and the origin of the Cartesian axes system is midway between the surfaces of the nearly touching spheres with $z = 0$ being the median plane (see Fig.~\ref{fig:bigsmallspheres}a).

In Fig.~\ref{fig:bigsmallspheres}b, we show the variation of the potential in the median plane obtained using the conventional boundary integral method (CBIM in Eq.~(\ref{eq:CBIMfield})) and by the present non-singular boundary regularized integral equation formulation (BRIEF in Eq.~(\ref{eq:BRIEFfield})). At this small separation ($a/h = 1000$, or $\kappa h = 10^{-3}$), the numerical precision of calculating the integral on one surface is adversely affected by the intrinsic near-singular behavior of the CBIM from the nearby surface is evident in the large errors or variations in the region $|x|/a < 0.5$. In contrast, the non-singular nature of the BRIEF means that in the same region, the calculation of the potential is not sensitive to the influence of proximal surfaces and that the potential variation is smooth and well-behaved as expected.

\subsubsection{Forces and torques}

The force acting on a colloid particle is calculated by integrating the Maxwell stress tensor over the particle surface. In the Debye-H$\ddot{\text{u}}$ckel model, the $ij$-component of the Maxwell stress tensor, $\sigma_{ij}$, is \cite{bell1970}
\begin{align}
  \sigma_{ij} = \epsilon_{0} \epsilon_{out} \left [ {E}_{i} {E}_{j} - \frac{1}{2} \left[ E_{k}E_{k} + \left(\kappa \phi \right)^2 \right] \delta_{ij} \right ]
\end{align}
where $E_{i}$ is the $i$th component of electric field $\boldsymbol{E}$, and $\delta_{ij}$ is the Kronecker Delta function. Thus, the $i$th component of the force, $\boldsymbol{F}$, acting on the particle is
\begin{align}
F_{i} = \int_{S} \sigma_{ij} n_{j} \text{ d}S
\end{align}
in which $n_{j}$ is the $j$th component of the outward surface normal. The torque about the Cartesian axis $i$ is calculated using
\begin{align}
  T_{i} = \int_{S} \varepsilon_{ijk} r_{j} \left(\sigma_{km} n_{m}\right) \text{ d}S,
\end{align}
where $\varepsilon_{ijk}$ is the three-dimensional Levi-Civita symbol, and $r_{j}$ is the $j$th component of the vector between the surface node and the center of the particle. For a pair of particles we confirm that the numerical results for the force and torque are consistent whether we integrate over the surface of one particle or the other.

\begin{figure}  
  \centering
  \subfloat[]{ \includegraphics[width=0.45\textwidth]{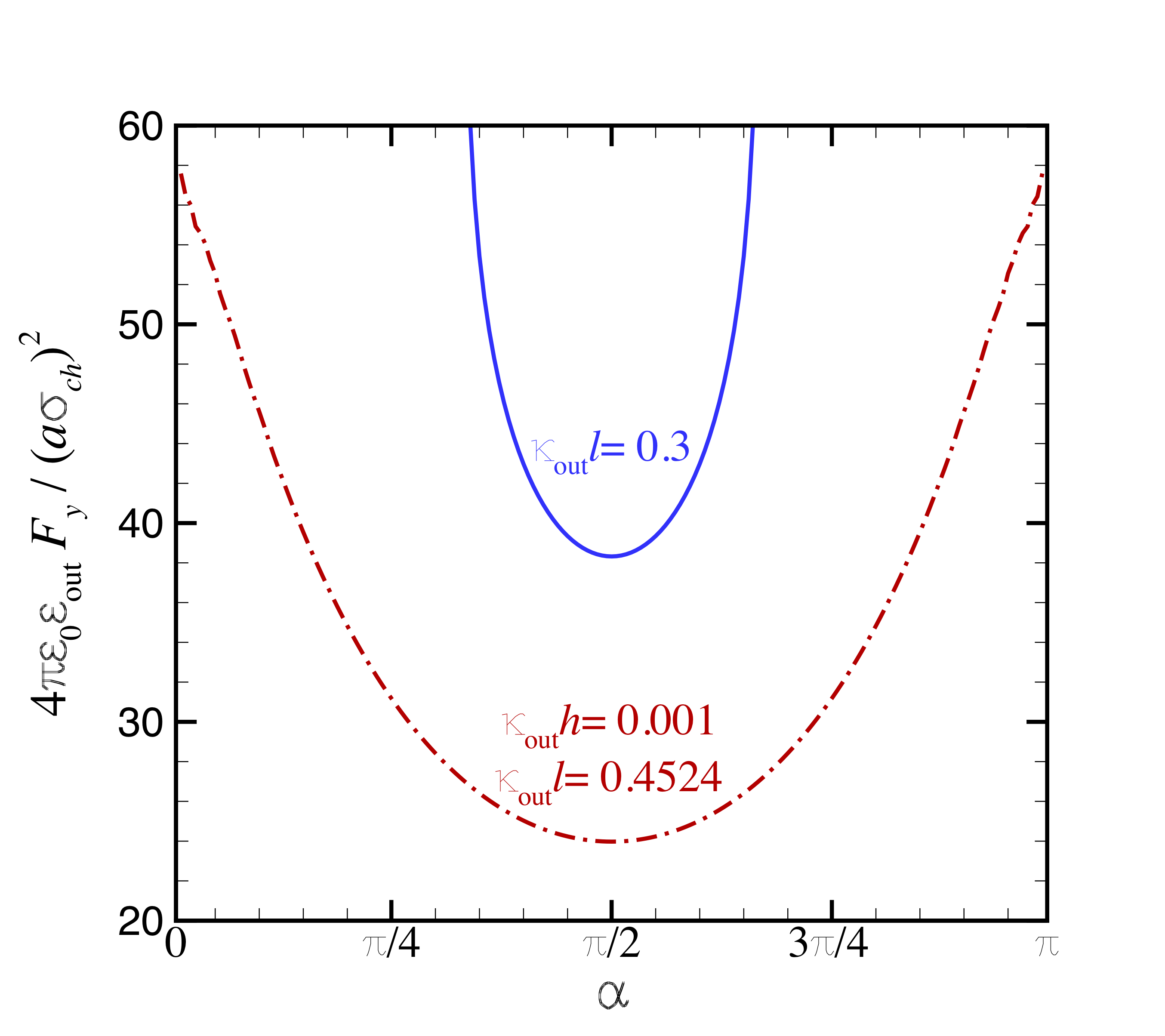} }
  \subfloat[]{ \includegraphics[width=0.45\textwidth]{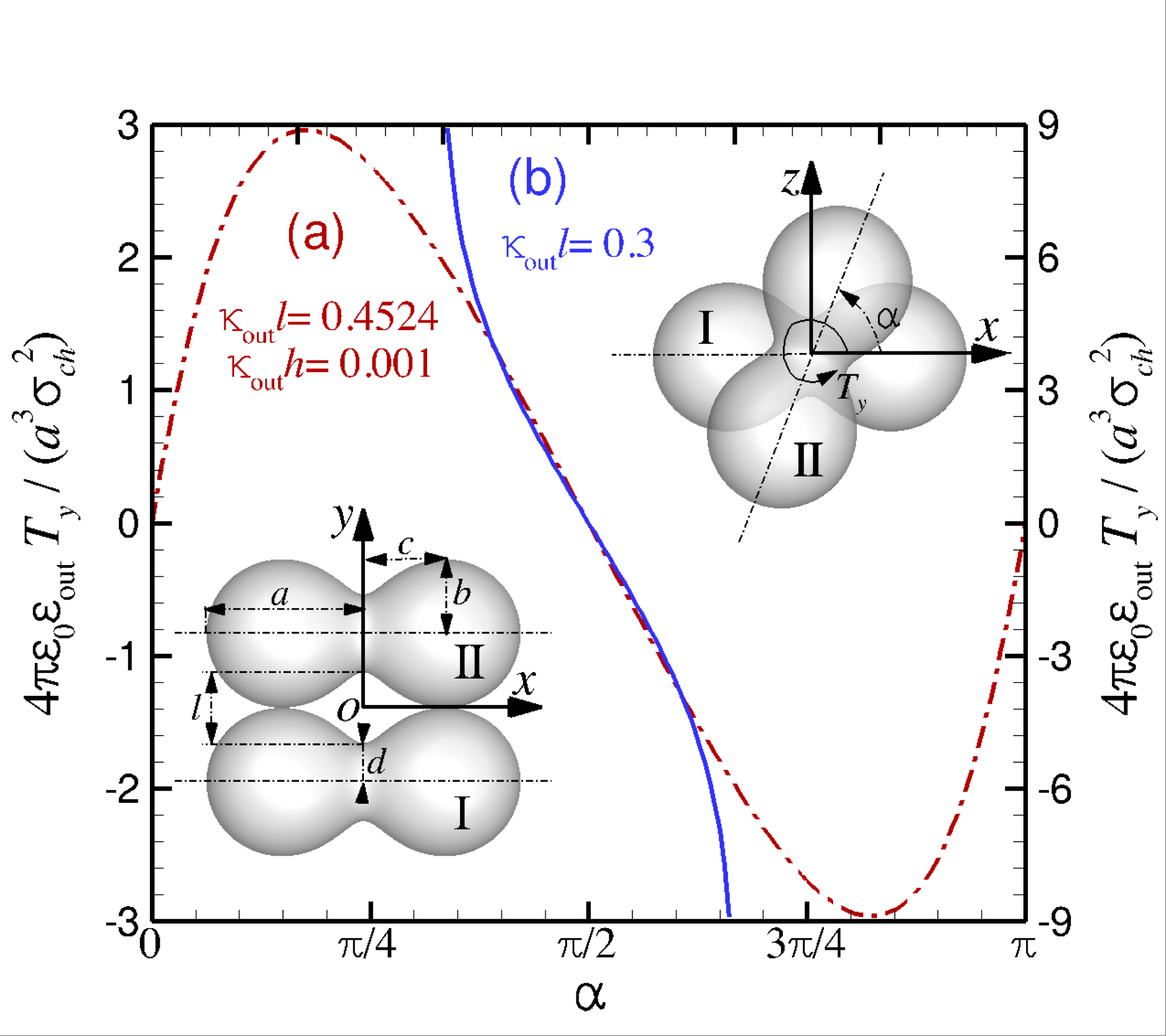} }
\caption{\label{fig:dumbbells_force_torque} (Color) (a) The force between two dumbbell particles as a function of the relative orientation at separation $\kappa l = 0.4524$ (see inset for definition of $l$). This corresponds to a small distance of closest approach, $\kappa h = 10^{-3}$ between the dumbbells. At separation $\kappa l = 0.3$ the narrow waists of the two dumbbells fit around each other thereby restricting the relative orientation to be only within a limited range about $\pi/2$. (b) The corresponding torque about the $y$-axis. Each dumbbell surface is represented by 1200 quadratic elements with 2402 nodes.}
\end{figure}

In Fig.~\ref{fig:dumbbells_force_torque} we show results for the force and torque between two identically charged dumbbells, described by Eq.~(\ref{eq:dumbbell}), that carry a uniform surface charge density, $\sigma_{ch}$. The dumbbells have dielectric constant $\epsilon_{in} = 2$ and are immersed in a solvent of dielectric constant $\epsilon_{out} = 80$ and Debye length $1/\kappa = 7.955$ {\AA}. The relevant dimensionless parameters are: $\kappa a = 1$, $b/a=0.4757$, $c/a=0.5303$, $d/a=0.25$. We give results for two separations: (i) at $\kappa l = 0.4524$ (see Fig. \ref{fig:dumbbells_force_torque}b for the definition) so that the distance of closest approach between surfaces of the dumbbells at their widest part is very small: $\kappa h = 10^{-3}$; and (ii) at $\kappa l = 0.3$, where the waists of the 2 dumbbells dovetail around each other so that the relative orientation between the axes of the dumbbells is confined within a limited range around $\pi/2$. These results show that present non-singular boundary regularized integral equation formulation (BRIEF) is very robust and being able to furnish stable numerical results when the system parameters such as separation and size ratios are at quite extreme limits. 

\section{Conclusions}
The boundary integral method is a very powerful approach for studying molecular and colloidal electrostatics. Its principal advantage from the physical point of view is that the solution is cast in terms of solving for unknowns on the surfaces that define the charged system. This is particularly beneficial if the surfaces have complex and important facets or when the primary interest is in the potentials and electric fields near surfaces or forces at small separations. Unlike volume based methods such as finite difference or finite element methods, one does not have to be concerned with multi-scale meshing issues of the 3D domain or with the behavior of the potential and field at infinity.

The technical challenge of boundary integral methods in giving rise to dense matrix systems that required impractical execution times has been overcome with the development of fast algorithms that are of very acceptable $O(N \log N)$ complexity~\cite{bardhan2007}. The remaining hurdle facing the wider adoption of the boundary integral method is the appearance of mathematical singularities in the conventional boundary integral method. While these singularities have no physical origin, they make it very difficult to use higher order methods to represent surfaces more accurately or to develop algorithms to evaluate the surface integrals with higher precision.

In this paper we have focused on a general formulation of the boundary integral method for the Debye-H$\ddot{\text{u}}$ckel model that removes the singular behavior that has to date been accepted as an intrinsic part of the conventional boundary integral equation method. By re-casting the problem in a way that better reflects the non-singular physical behavior of the system, we have shown with examples drawn from molecular and colloidal electrostatics that this is a robust, efficient and accurate approach. The removal of physically irrelevant singularities affords considerable savings in coding effort and results in orders of magnitude improvement in numerical precision for the same problem size.  The enhanced accuracy also allows the electric field at boundaries to be calculated easily and accurately without having to solve hypersingular integral equations. As a consequence, physically important quantities such as forces and torques can be calculated easily. These advances should therefore provide an impetus to use the present non-singular boundary regularized integral equation formulation (BRIEF) to tackle complex and important problems. The present framework can easily be accommodated in existing boundary integral codes with the addition of a few lines of new code whereas all existing code that handles the singularities can be discarded.

\begin{acknowledgments}
This work is supported by the Australian Research Council (ARC) through a Discovery Early Career Research Award to QS and a Discovery Project Grant to DYCC.
\end{acknowledgments}

\appendix

\section{Kirkwood ion potential distribution}\label{appx:kirkwoodsol}
By positioning the point ion of charge, $q$, on the $z$-axis at $\boldsymbol{x}_s = (0, 0, r_s)$, the electrostatic potential, $\phi(\boldsymbol{x})$, inside and outside a Kirkwood ion with a Stern layer is governed by Eqs.~(\ref{eq:phi_de_a}) - (\ref{eq:phi_de_c}). The analytic solution of which can be expressed as infinite series in Legendre polynomials, $P_n(\cos \theta)$, of order $n$ that depends on the polar angle $\theta$ measured relative to the $z$-axis \cite{kirkwood1934}
\begin{eqnarray} 
  \phi(\boldsymbol{x}) 
  &=&    \frac{q}{4 \pi \epsilon_0 \epsilon_{in}} \sum_{n=0}^\infty \frac{r_<^n}{r_>^{n+1}} P_n(cos \theta) + 
     \frac{q}{4 \pi \epsilon_0 \epsilon_{in} a} \sum_{n=0}^\infty A_n \frac{r^n}{a^n} P_n(\cos \theta), \quad 0 < r < a, \label{eq:phi_a} \\
  &=&   \frac{q}{4 \pi \epsilon_0 \epsilon_{st} a} \sum_{n=0}^\infty \left [ B_n \frac{r^n}{a^n} + C_n \frac{a^{n+1}}{r^{n+1}} \right ]P_n(\cos \theta), \qquad a < r < b,  \label{eq:phi_b} \\
  &=&   \frac{q}{4 \pi \epsilon_0 \epsilon_{out} a} \sum_{n=0}^\infty D_n k_n(\kappa r)  P_n(\cos \theta), \qquad  r > b. \label{eq:phi_c}
  \end{eqnarray}
The first sum in Eq.~(\ref{eq:phi_a}) is the spherical harmonic expansion of the coulomb potential due to the point charge at $\boldsymbol{x}_s$ with $r_> \equiv$ max$(r, r_s)$ and $r_<  \equiv$ min$(r, r_s)$. The second sum in Eq.~(\ref{eq:phi_a}) is the reaction potential, $\phi_{react}(\boldsymbol{x})$ -- see Eq.~(\ref{eq:phi_react}).  In Eq.~(\ref{eq:phi_c}), $k_n(z)$ is a modified spherical function of the second kind of order $n$ \cite{abramowitz1970}. If the solvent in the region $r > b$ is a dielectric, $k_n(\kappa r)$ is to be replaced by $(b/r)^{n+1}$. The 4 coefficients $A_n, ..., D_n$ are determined by the continuity of $\phi$ and $\epsilon (\partial \phi / \partial r)$ at $r = a$ and $r = b$.

The continuity of $\phi$ and $\epsilon (\partial \phi / \partial r)$ at $r = a$ gives
\begin{eqnarray}
  \frac{1}{\epsilon_{in}} \left (\frac{r_s}{a} \right )^n + \frac{1}{\epsilon_{in}} A_n &=& \frac{1}{\epsilon_{st}} (B_n + C_n) \\
  -(n+1) \left (\frac{r_s}{a} \right )^n + n A_n &=& n B_n - (n+1) C_n,
\end{eqnarray}
and at at $r = b$ gives
\begin{eqnarray}
  \frac{1}{\epsilon_{st}} \left [   B_n \left (\frac{b}{a} \right )^n + C_n \left (\frac{a}{b} \right )^{n+1} \right] &=& \frac{1}{\epsilon_{out}} k_n(\kappa b) D_n \\
  n B_n \left (\frac{b}{a} \right )^n - (n+1) C_n \left (\frac{a}{b} \right )^{n+1} &=& (\kappa b) k'_n(\kappa b) D_n.
  \end{eqnarray}
These  4 equations can readily be solved for $A_n$. For a dielectric solvent with $\kappa = 0$, we make the replacements: $k_n(\kappa b) D_n \rightarrow D_n$ and $(\kappa b) k'_n(\kappa b) D_n \rightarrow -(n+1) D_n$.

From Eqs. (\ref{eq:phi_react}) and (\ref{eq:E_solv}), the solvation energy, $E_{solv}$ can be expressed in terms of the reaction potential, $\phi_{react}(\boldsymbol{x})$ evaluated at the point ion: $(r, \theta) = (r_s, 0)$ \cite{kirkwood1934}
\begin{align}
  E_{solv} = \frac{1}{2} q \left ( \frac{q}{4 \pi \epsilon_0 \epsilon_{in} a} \sum_{n=0}^\infty A_n \frac{r_s^n}{a^n}\right).
\end{align}

\section{Numerical implementation}\label{appx:ni}
As an example of how our boundary regularized integral equation formulation (BRIEF) given by Eq. (\ref{eq:BRIEF}) can be discretized to give a linear system to be solved for the unknown potential, we consider the Kirkwood ion defined by Eqs. (\ref{eq:phi_de_a}) - (\ref{eq:phi_de_c}). This simple model has all the physical features of general problems in molecular electrostatics in which a charged molecule is modeled as having point ions embedded in a dielectric region of dielectric constant, $\epsilon_{in}$, that is immersed in an electrolyte of dielectric constant, $\epsilon_{out}$, and ionic concentration characterized by a Debye length, $1/\kappa$. In addition, a thin dielectric Stern layer of dielectric constant, $\epsilon_{st}$, that excludes ionic species separates the electrolyte from the charged molecule.

\begin{figure}[!t]
\centering{}\includegraphics[width=0.75\textwidth]{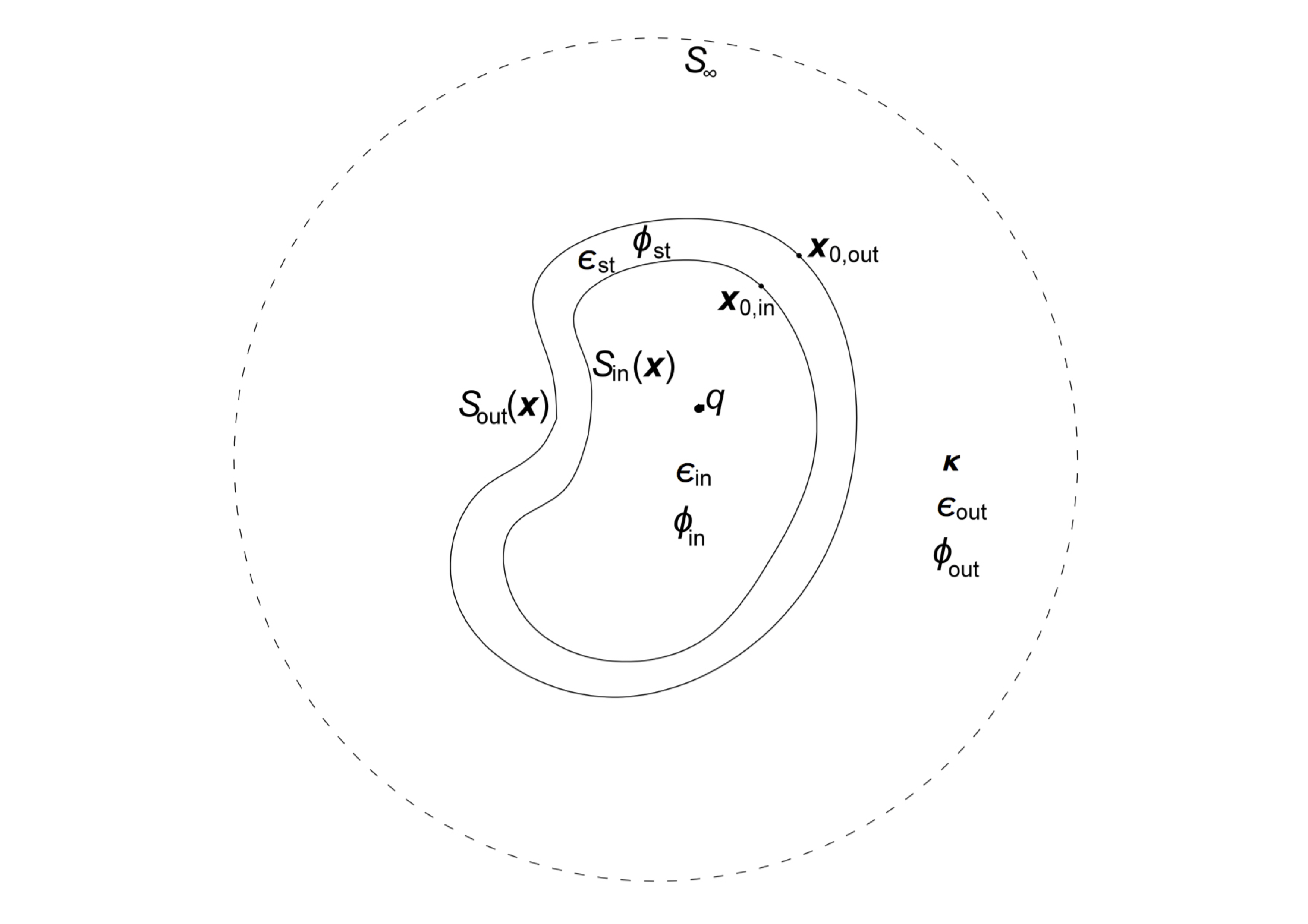} \caption{\label{Fig:ni}  Schematic sketch of the physical problem simulated by the boundary regularized integral equation formulation (BRIEF).}
\end{figure}

Such a system comprises two boundaries that are the inner, $S_{in}$, and outer, $S_{out}$, surfaces of the Stern layer (see Fig. \ref{Fig:ni}). The potential is therefore defined by implementing the BRIEF, Eq. (\ref{eq:BRIEF}), at these two surface that result in a pair of coupled surface integral equations. The usual electrostatic boundary conditions for the continuity of $\phi$ and $(\epsilon \; \partial{\phi} / \partial{n})$ are applied at these two surfaces. On the inner surface,  $S_{in}$, we have 
\begin{align}
  \phi_{st, in} & = \phi_{in}\; \label{eq:bcin}\\
  \nonumber\\
  \frac{ \partial{ \phi_{st, in} } }{ \partial{n} } & = \frac{ \epsilon_{in} }{ \epsilon_{st} } \frac{ \partial{ \phi_{in} } }{ \partial{n} } \label{eq:bcinn}
\end{align}
where $\phi_{in}$ and $\partial{ \phi_{in} } / \partial{n}$ are the potential and its normal derivative within the particle that are on $S_{in}$ and  $\phi_{st, in}$ and $\partial{ \phi_{st, in} } / \partial{n}$ are the potential and its normal derivative within the Stern layer that are on $S_{in}$. Similarly, on the outer surface, $S_{out}$, we have 
\begin{align}
  \phi_{st, out} & = \phi_{out}\;  \label{eq:bcou}\\
  \nonumber\\
  \frac{ \partial{ \phi_{st, out} } }{ \partial{n} } & = \frac{ \epsilon_{out} }{ \epsilon_{st} } \frac{ \partial{ \phi_{out} } }{ \partial{n} } \label{eq:bcoutn} 
\end{align}
where  $\phi_{out}$ and $\partial{ \phi_{out} } / \partial{n}$ are the potential and its normal derivative within the particle that are on $S_{out}$ and  $\phi_{st, out}$ and $\partial{ \phi_{st, out} } / \partial{n}$ are the potential and its normal derivative within the Stern layer that are on $S_{out}$.

Using Eq. (\ref{eq:BRIEF}) for the potential distribution in the region prescribed by Eq. (\ref{eq:phi_de_a}), we can write a relation between the potential $\phi_{in}$ and its normal derivative $\partial{ \phi_{in} } / \partial{ n }$ on $S_{in}$ that is the surface of the molecule or the inner surface of the Stern layer as
\begin{align}\label{eq:phiinBRIEF}
&\quad \int_{S_{in}}  \left[  \phi_{in}(\boldsymbol{x}) \frac{\partial{G_{0}}}{\partial{n}} - \phi_{in}(\boldsymbol{x}_0) g(\boldsymbol{x}) \frac{\partial{G_0}}{\partial{n}}    + \phi_{in}(\boldsymbol{x}_0)  \frac{\partial{g(\boldsymbol{x})}}{\partial{n}}G_{0} \right]\text{ d}S(\boldsymbol{x}) \nonumber \\ 
&- \int_{S_{in}} \left[  \frac{\partial{\phi_{in}(\boldsymbol{x})}}{\partial{n}} G_{0} - \frac{\partial{\phi_{in}(\boldsymbol{x}_0)}}{\partial{n}}  \frac{\partial{f(\boldsymbol{x})}}{\partial{n}} G_0  +\frac{\partial{\phi_{in}(\boldsymbol{x}_0)}}{\partial{n}} f(\boldsymbol{x}) \frac{\partial{G_0}}{\partial{n}} \right] \text{ d}S(\boldsymbol{x})  \nonumber \\
& = \frac{ q} { 4 \pi \epsilon_0 \epsilon_{in} |\boldsymbol{x}_{0} - \boldsymbol{x}_s| }. 
\end{align}

For the potential in the Stern layer prescribed by Eq. (\ref{eq:phi_de_b}), the surface integrals of the BRIEF will be taken on both the inner surface, $S_{st, in}$, and the outer surface, $S_{st, out}$, of the Stern layer as
\begin{align}\label{eq:phistBRIEF}
\nonumber
&\quad \int_{S_{in}}  \left[  \phi_{in}(\boldsymbol{x}) \frac{\partial{G_{0}}}{\partial{n}} - \phi_{in}(\boldsymbol{x}_0) g(\boldsymbol{x}) \frac{\partial{G_0}}{\partial{n}}    + \phi_{in}(\boldsymbol{x}_0)  \frac{\partial{g(\boldsymbol{x})}}{\partial{n}}G_{0} \right]\text{ d}S(\boldsymbol{x}) \nonumber \\ 
&- \int_{S_{in}} \frac{ \epsilon_{in} }{ \epsilon_{st} }  \left[  \frac{\partial{\phi_{in}(\boldsymbol{x})}}{\partial{n}} G_{0}  - \frac{\partial{\phi_{in}(\boldsymbol{x}_0)}}{\partial{n}}  \frac{\partial{f(\boldsymbol{x})}}{\partial{n}} G_0  +\frac{\partial{\phi_{in}(\boldsymbol{x}_0)}}{\partial{n}} f(\boldsymbol{x}) \frac{\partial{G_0}}{\partial{n}} \right] \text{ d}S(\boldsymbol{x})  \nonumber \\
&+ \int_{S_{out}}  \left[  \phi_{out}(\boldsymbol{x}) \frac{\partial{G_{0}}}{\partial{n}} - \phi_{out}(\boldsymbol{x}_0) g(\boldsymbol{x}) \frac{\partial{G_0}}{\partial{n}}    + \phi_{out}(\boldsymbol{x}_0)  \frac{\partial{g(\boldsymbol{x})}}{\partial{n}}G_{0} \right]\text{ d}S(\boldsymbol{x}) \nonumber \\ 
&- \int_{S_{out}} \frac{ \epsilon_{out} }{ \epsilon_{st} }  \left[  \frac{\partial{\phi_{out}(\boldsymbol{x})}}{\partial{n}} G_{0} - \frac{\partial{\phi_{out}(\boldsymbol{x}_0)}}{\partial{n}}  \frac{\partial{f(\boldsymbol{x})}}{\partial{n}} G_0  +\frac{\partial{\phi_{out}(\boldsymbol{x}_0)}}{\partial{n}} f(\boldsymbol{x}) \frac{\partial{G_0}}{\partial{n}} \right] \text{ d}S(\boldsymbol{x})  \nonumber \\
& = 0,
\end{align}
where Eqs. (\ref{eq:bcin}) to (\ref{eq:bcoutn}) have been used. 

Finally, the integrals from the BRIEF for the potential in the external domain are
\begin{align} \label{eq:finalBRIEF}
&\quad \int_{S_{out}}  \left[  \phi_{out}(\boldsymbol{x}) \frac{\partial{ G}}{\partial{n}} - \phi_{out}(\boldsymbol{x}_0) g(\boldsymbol{x}) \frac{\partial{G_0}}{\partial{n}}    + \phi_{out}(\boldsymbol{x}_0)  \frac{\partial{g(\boldsymbol{x})}}{\partial{n}}G_{0} \right]\text{ d}S(\boldsymbol{x}) \nonumber \\ 
&- \int_{S_{out}} \left[  \frac{\partial{\phi_{out}(\boldsymbol{x})}}{\partial{n}} G  - \frac{\partial{\phi_{out}(\boldsymbol{x}_0)}}{\partial{n}}  \frac{\partial{f(\boldsymbol{x})}}{\partial{n}} G_0 +\frac{\partial{\phi_{out}(\boldsymbol{x}_0)}}{\partial{n}} f(\boldsymbol{x}) \frac{\partial{G_0}}{\partial{n}} \right] \text{ d}S(\boldsymbol{x})  \nonumber \\
&+ \int_{S_{\infty}}  \left[  \phi_{out}(\boldsymbol{x}) \frac{\partial{ G}}{\partial{n}} - \phi_{out}(\boldsymbol{x}_0) g(\boldsymbol{x}) \frac{\partial{G_0}}{\partial{n}}    + \phi_{out}(\boldsymbol{x}_0)  \frac{\partial{g(\boldsymbol{x})}}{\partial{n}}G_{0} \right]\text{ d}S(\boldsymbol{x}) \nonumber \\ 
&- \int_{S_{\infty}} \left[  \frac{\partial{\phi_{out}(\boldsymbol{x})}}{\partial{n}} G  - \frac{\partial{\phi_{out}(\boldsymbol{x}_0)}}{\partial{n}}  \frac{\partial{f(\boldsymbol{x})}}{\partial{n}} G_0  +\frac{\partial{\phi_{out}(\boldsymbol{x}_0)}}{\partial{n}} f(\boldsymbol{x}) \frac{\partial{G_0}}{\partial{n}} \right] \text{ d}S(\boldsymbol{x})  \nonumber \\
& = 0 \nonumber  \\
\end{align}
The integrals over the surface at infinity, $S_{\infty}$ in the last two terms in Eq. (\ref{eq:finalBRIEF}) can be easily simplified to $4\pi \phi_{out}(\boldsymbol{x}_0) $ when $g(\boldsymbol{x})$ and $f(\boldsymbol{x})$ are chosen according to Eq. (\ref{eq:simpleFandG}) thus giving
\begin{align}\label{eq:phioutBRIEF}
&\quad 4\pi \phi_{out}(\boldsymbol{x}_0) +  \int_{S_{out}}  \left[  \phi_{out}(\boldsymbol{x}) \frac{\partial{ G}}{\partial{n}} - \phi_{out}(\boldsymbol{x}_0) g(\boldsymbol{x}) \frac{\partial{G_0}}{\partial{n}}    + \phi_{out}(\boldsymbol{x}_0)  \frac{\partial{g(\boldsymbol{x})}}{\partial{n}}G_{0} \right]\text{ d}S(\boldsymbol{x}) \nonumber \\ 
&- \int_{S_{out}} \left[  \frac{\partial{\phi_{out}(\boldsymbol{x})}}{\partial{n}} G  - \frac{\partial{\phi_{out}(\boldsymbol{x}_0)}}{\partial{n}}  \frac{\partial{f(\boldsymbol{x})}}{\partial{n}} G_0 +\frac{\partial{\phi_{out}(\boldsymbol{x}_0)}}{\partial{n}} f(\boldsymbol{x}) \frac{\partial{G_0}}{\partial{n}} \right] \text{ d}S(\boldsymbol{x})  \nonumber \\
& = 0. 
\end{align}
In Eqs. (\ref{eq:phiinBRIEF}) to (\ref{eq:phioutBRIEF}), $G \equiv  G(\boldsymbol{x},\boldsymbol{x}_{0})$ is given in Eq. (\ref{eq:GREEN}) and $G_{0} \equiv G_{0}(\boldsymbol{x},\boldsymbol{x}_{0})$ is given in Eq. (\ref{eq:LaplaceGREEN}).

\begin{figure}[!t]
\centering{}\includegraphics[width=0.75\textwidth]{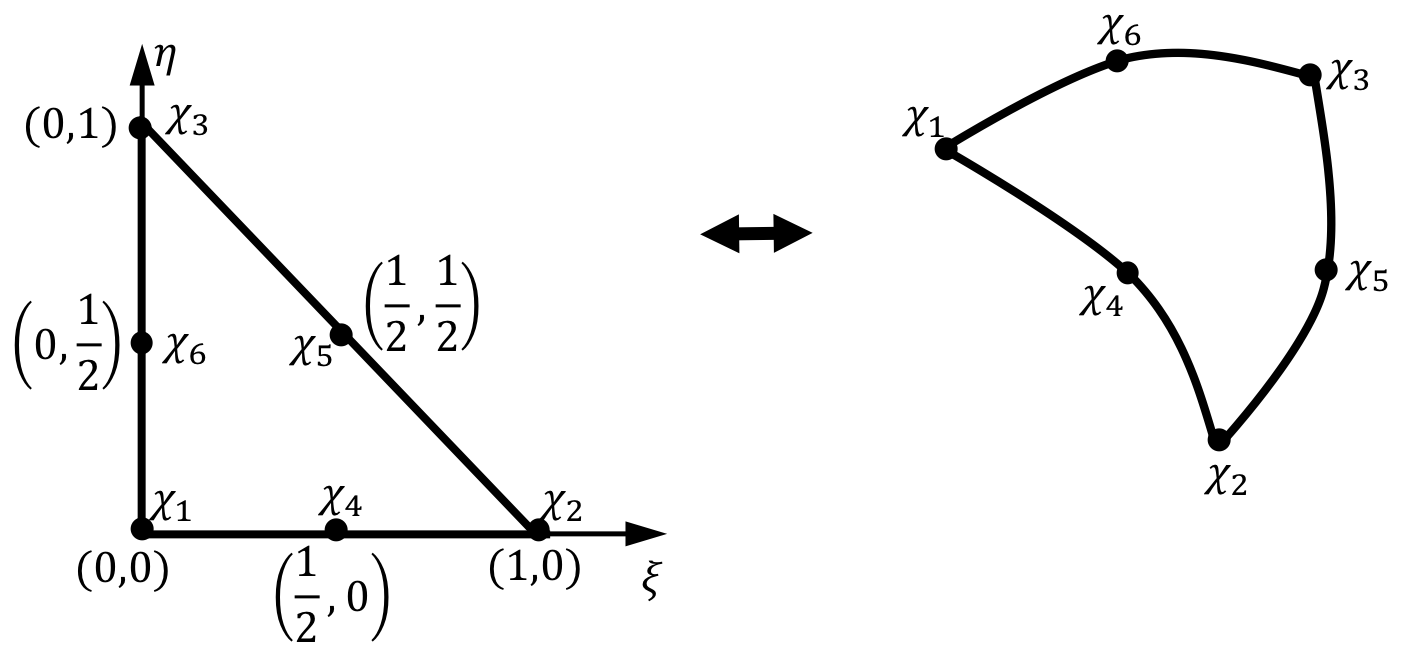} \caption{\label{Fig:quadratic}  The  interpolation scheme on a quadratic surface element in the local surface variables ($\xi$, $\eta$).}
\end{figure}

To solve Eqs. (\ref{eq:phiinBRIEF}) to (\ref{eq:phioutBRIEF}) to obtain $\phi_{in}$ $\phi_{out}$, $\partial{\phi_{in}}/\partial{n}$ and $\partial{\phi_{out}}/\partial{n}$ numerically, the surfaces, $S_{in}$ and $S_{out}$, are discretized by using quadratic triangular area elements on which each element is bounded by 3 nodes on the vertices and 3 nodes on the edges, see Fig. \ref{Fig:quadratic} for a total of $N$ nodes on the surface. The coordinates of a point in each element and the function values there are obtained by quadratic interpolation from the values at the nodes using the standard quadratic interpolation function ($\nu \equiv 1 - \xi - \eta$)
\begin{align}
  \chi = \; & \nu(2\nu -1) \; \chi_{1} + \xi(2\xi -1) \; \chi_{2}+\eta(2\eta -1) \; \chi_{3} \nonumber \\
  &  +4 \nu\xi  \; \chi_{4} +4\xi \eta  \; \chi_{5} +4 \eta \nu  \; \chi_{6},
\end{align}
in terms of the local coordinates $(\xi, \eta)$ (see Fig. \ref{Fig:quadratic}).

The solution of the potentials and their normal derivatives on the surfaces are expressed in terms of the values at the $N$ surface nodes. As illustrated in Fig. \ref{Fig:ni}, when two observation points, $\boldsymbol{x}_ { {0}_{in} }$ and $\boldsymbol{x}_ { {0}_{out} }$, are chosen to be located on the inner surface, $S_{in}$, and outer surface, $S_{out}$, of the Stern layer, respectively, the surface integral equations, Eqs. (\ref{eq:phiinBRIEF}) to (\ref{eq:phioutBRIEF}), can be expressed as a system of linear equations 
\begin{align}
 & {\cal{H}}_{in}^{0}(\boldsymbol{x}_{0_{in}}) \cdot \phi_{in} - {\cal{G}}_{in}^{0}(\boldsymbol{x}_{0_{in}}) \cdot \frac{ \partial{\phi_{in}} }{ \partial{n} } = \frac{ q } { 4 \pi \epsilon_0 \epsilon_{in} |\boldsymbol{x}_ { {0}_{in} } - \boldsymbol{x}_s|  } ,\\
 & {\cal{H}}_{in}^{0}(\boldsymbol{x}_{0_{in}}) \cdot \phi_{in} - \frac{ \epsilon_{in} }{ \epsilon_{st} }  {\cal{G}}_{in}^{0}(\boldsymbol{x}_{0_{in}}) \cdot \frac{ \partial{\phi_{in}} }{ \partial{n} } + {\cal{H}}_{out}^{0}(\boldsymbol{x}_{0_{in}}) \cdot \phi_{out} - \frac{ \epsilon_{out} }{ \epsilon_{st} }  {\cal{G}}_{out}^{0}(\boldsymbol{x}_{0_{in}}) \cdot \frac{ \partial{\phi_{out}} }{ \partial{n} }  = 0 ,\\
  & {\cal{H}}_{in}^{0}(\boldsymbol{x}_{0_{out}}) \cdot \phi_{in} - \frac{ \epsilon_{in} }{ \epsilon_{st} }  {\cal{G}}_{in}^{0}(\boldsymbol{x}_{0_{out}}) \cdot \frac{ \partial{\phi_{in}} }{ \partial{n} } + {\cal{H}}_{out}^{0}(\boldsymbol{x}_{0_{out}}) \cdot \phi_{out} - \frac{ \epsilon_{out} }{ \epsilon_{st} }  {\cal{G}}_{out}^{0}(\boldsymbol{x}_{0_{out}}) \cdot \frac{ \partial{\phi_{out}} }{ \partial{n} }  = 0 ,\\
 & \left[ 4\pi \boldsymbol{I} + {\cal{H}}_{out}(\boldsymbol{x}_{0_{out}}) \right] \cdot \phi_{out} - {\cal{G}}_{out}(\boldsymbol{x}_{0_{out}}) \cdot \frac{ \partial{\phi_{out}} }{ \partial{n} }  = 0,
\end{align}
where $\boldsymbol{I}$ is the identity matrix, the elements of the matrices ${\cal{H}}_{in}^{0}(\boldsymbol{x}_{0_{in}})$, ${\cal{H}}_{out}^{0}(\boldsymbol{x}_{0_{in}})$, ${\cal{G}}_{in}^{0}(\boldsymbol{x}_{0_{in}})$ and ${\cal{G}}_{out}^{0}(\boldsymbol{x}_{0_{in}})$ are the results of integrals (influence matrices) over the surface elements involving the unknown $4N$-vector ($\phi_{in}$ $\phi_{out}$, $\partial{\phi_{in}}/\partial{n}$, $\partial{\phi_{out}}/\partial{n}$) corresponding to Eqs. (\ref{eq:phiinBRIEF}) and (\ref{eq:phistBRIEF}) as $\boldsymbol{x}_ { {0}_{in} }$ is on $S_{in}$, and  ${\cal{H}}_{in}^{0}(\boldsymbol{x}_{0_{out}})$, ${\cal{H}}_{out}^{0}(\boldsymbol{x}_{0_{out}})$, ${\cal{H}}_{out}(\boldsymbol{x}_{0_{out}})$, ${\cal{G}}_{in}^{0}(\boldsymbol{x}_{0_{out}})$, ${\cal{G}}_{out}^{0}(\boldsymbol{x}_{0_{out}})$ and ${\cal{G}}_{out}(\boldsymbol{x}_{0_{out}})$ are the results of integrals over the surface elements corresponding to Eqs. (\ref{eq:phistBRIEF}) and (\ref{eq:phioutBRIEF}) as $\boldsymbol{x}_ { {0}_{out} }$ is on $S_{out}$. Since the surface integral equations (\ref{eq:phiinBRIEF}) to (\ref{eq:phioutBRIEF}) do not have any singular behavior, these matrix elements can be calculated accurately using standard Gauss quadrature. The above set of equations is a $4N\times 4N$ linear system for the unknown complex $4N$-vectors: $\phi_{in}$ $\phi_{out}$, $\partial{\phi_{in}}/\partial{n}$, $\partial{\phi_{out}}/\partial{n}$ on the surface in the final form
\begin{align}
 & \begin{bmatrix}
    {\cal{H}}_{in}^{0}(\boldsymbol{x}_{0_{in}}) & \boldsymbol{0} & - {\cal{G}}_{in}^{0}(\boldsymbol{x}_{0_{in}})  & \boldsymbol{0} \\
    {\cal{H}}_{in}^{0}(\boldsymbol{x}_{0_{in}}) & {\cal{H}}_{out}^{0}(\boldsymbol{x}_{0_{in}})  & -( \epsilon_{in} / \epsilon_{st} ){\cal{G}}_{in}^{0}(\boldsymbol{x}_{0_{in}})  &  -( \epsilon_{out} / \epsilon_{st} ) {\cal{G}}_{out}^{0}(\boldsymbol{x}_{0_{in}})  \\
    {\cal{H}}_{in}^{0}(\boldsymbol{x}_{0_{out}}) & {\cal{H}}_{out}^{0}(\boldsymbol{x}_{0_{out}})  & -( \epsilon_{in} / \epsilon_{st} ){\cal{G}}_{in}^{0}(\boldsymbol{x}_{0_{out}})  &  -( \epsilon_{out} / \epsilon_{st} ) {\cal{G}}_{out}^{0}(\boldsymbol{x}_{0_{out}})  \\ 
    \boldsymbol{0}  & \left[ 4\pi \boldsymbol{I} + {\cal{H}}_{out}(\boldsymbol{x}_{0_{out}})  \right] & \boldsymbol{0}  & -{\cal{G}}_{out}(\boldsymbol{x}_{0_{out}}) 
  \end{bmatrix}
    \left[ \begin{array}{c} \phi_{in} \\ \phi_{out} \\ \partial{\phi_{in}}/\partial{n} \\ \partial{\phi_{out}}/\partial{n} \end{array} \right] \nonumber \\
 & \hspace{5.5cm} =
 \left[ \begin{array}{c}  q /( 4 \pi \epsilon_0 \epsilon_{in} |\boldsymbol{x}_ { {0}_{in} } - \boldsymbol{x}_{s}|  ) \\ \boldsymbol{0} \\ \boldsymbol{0} \\ \boldsymbol{0} \end{array} \right].
\end{align}

\section{Calculating the potential and field}\label{appx:Efield}
 
\begin{figure}[!t]
\centering
\includegraphics[width=0.42\textwidth] {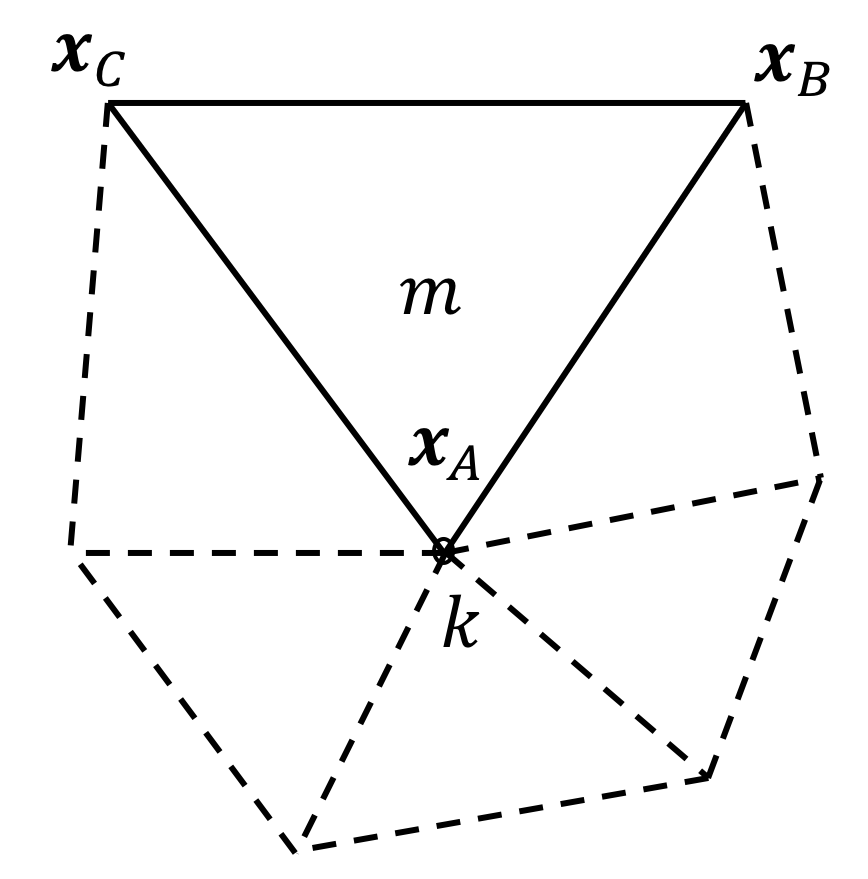}
\caption{Calculating the electric field at node $k$ by averaging over all area elements that share the node.}
\label{fig:elmntkm}
\end{figure}

 The absence of singular integrals in the BRIEF means that the potential on the surface, $S$, can be evaluated accurately without numerical instabilities. Consequently, the electric field on the surface can be calculated without the need to solve hypersingular integral equations \cite{lu2005}. Suppose we seek the electric field, $\boldsymbol{E} = - \nabla \phi$ at node $k$ on the mesh that represents the surface $S$, shown in Fig.~\ref{fig:elmntkm}. Consider one of the surface elements, $m$, assumed for simplicity to be a planar triangle with vertices at $\boldsymbol{x}_A$ (node $k$), $\boldsymbol{x}_B$ and $\boldsymbol{x}_C$.  The normal component of the electric field of node $k$ at $\boldsymbol{x}_A$, estimated from element $m$ is $\boldsymbol{E}_{k,m}$  is given by $\partial \phi / \partial n$ from the boundary integral equation, and tangential components of $\boldsymbol{E}_{k,m}$ can be estimated \emph{via} finite differencing using the potential values at the nodes of this area element $m$

\begin{align} \label{eq:calcE}
\left\{\begin{aligned}
(\boldsymbol{x}_{A}-\boldsymbol{x}_{B})\cdot(-\boldsymbol{E}_{k,m})&\approx \phi(\boldsymbol{x}_A)-\phi(\boldsymbol{x}_B),\\
(\boldsymbol{x}_{A}-\boldsymbol{x}_{C})\cdot(-\boldsymbol{E}_{k,m})&\approx \phi(\boldsymbol{x}_A)-\phi(\boldsymbol{x}_C),\\
\boldsymbol{n}_{k}\cdot(-\boldsymbol{E}_{k,m})&\approx \left(\frac{\partial{\phi}}{\partial{n}}\right)_{k}.
\end{aligned}
\right.
\end{align}
Now all components of $\boldsymbol{E}_{k,m}$ can be found by solving Eq.~(\ref{eq:calcE}). However, all area elements that share node $k$ also contribute to the estimate of the electric field of node $k$ at $\boldsymbol{x}_A$, therefore the field $\boldsymbol{E}_k$ at node $k$ will be the weighted contribution from $N_k$ such elements according to
\begin{align} \label{eq:electricfield}
  \boldsymbol{E}_k = \sum_{m=1}^{N_k} w_m ~ \boldsymbol{E}_{k,m}.
\end{align}
The weight, $w_m$, corresponding to element $m$ is taken to be inversely proportional to its area, $S_m$: 
\begin{align}
  w_m = \frac {1/S_m} {\sum_{n=1}^{N_k} 1/S_n}.
\end{align}

 Also, BRIEF provides a robust way to calculate the potentials at field positions with the same level of accuracy within the entire solution domain.  In CBIM, the loss of accuracy due to the near singularity when the field position is close to the boundaries is usually more difficult to deal with compared to the singular behavior on the boundaries. 
 
 To calculate the potential accurately at position $\boldsymbol{x}_{p}$ in the 3D domain, we first use the CBIM to get $\phi(\boldsymbol{x}_{p})$, with $c_0=4\pi$,
 \begin{align}\label{eq:CBIMfield}
4\pi \phi(\boldsymbol{x}_{p}) + \int_{S} \phi(\boldsymbol{x}) \frac{\partial{G(\boldsymbol{x},\boldsymbol{x}_{p})}}{\partial{n}}\text{ d}S(\boldsymbol{x}) = \int_{S} G(\boldsymbol{x},\boldsymbol{x}_{p}) \frac{\partial{\phi(\boldsymbol{x})}}{\partial{n}}\text{ d}S(\boldsymbol{x}).
\end{align}
In the same manner, from Eq. (\ref{eq:LaplaceCBIM}), we have
\begin{align}\label{eq:LaplaceCBIMfield}
4\pi\psi(\boldsymbol{x}_{p}) + \int_{S} \psi(\boldsymbol{x}) \frac{\partial{G_0(\boldsymbol{x},\boldsymbol{x}_{p})}}{\partial{n}}\text{ d}S(\boldsymbol{x}) = \int_{S} G_0(\boldsymbol{x},\boldsymbol{x}_{p}) \frac{\partial{\psi(\boldsymbol{x})}}{\partial{n}}\text{ d}S(\boldsymbol{x}),
\end{align}
in which the $\boldsymbol{x}_{0}$ used to construct $\psi(\boldsymbol{x})$ in Eq. (\ref{eq:psi0fg}) is chosen as the node position that is closest to $\boldsymbol{x}_{p}$.
Subtracting Eq. (\ref{eq:LaplaceCBIMfield}) from Eq. (\ref{eq:CBIMfield}) yields
\begin{align}\label{eq:BRIEFfieldpre}
  \nonumber
4\pi \phi(\boldsymbol{x}_{p}) = 4\pi\psi(\boldsymbol{x}_{p})  & - \int_{S} \left[ \phi(\boldsymbol{x}) \frac{\partial{G(\boldsymbol{x},\boldsymbol{x}_{p})}}{\partial{n}} - \psi(\boldsymbol{x}) \frac{\partial{G_0(\boldsymbol{x},\boldsymbol{x}_{p})}}{\partial{n}} \right] \text{ d}S(\boldsymbol{x})   \\ 
& + \int_{S} \left[ G(\boldsymbol{x},\boldsymbol{x}_{p}) \frac{\partial{\phi(\boldsymbol{x})}}{\partial{n}} - G_0(\boldsymbol{x},\boldsymbol{x}_{p}) \frac{\partial{\psi(\boldsymbol{x})}}{\partial{n}} \right] \text{ d}S(\boldsymbol{x}).
\end{align}
The near singular behavior when $\boldsymbol{x}_{p}$ is close to the boundary can be eliminated by subtracting the BRIEF in Eq. (\ref{eq:BRIEF}) from the above boundary integral equation (\ref{eq:BRIEFfieldpre}). This then provides a numerically robust expression for $\phi(\boldsymbol{x}_{p})$ whose accuracy is not affected by the distance between $\boldsymbol{x}_{p}$ and any boundary:
\begin{align}\label{eq:BRIEFfield}
  \nonumber 
4\pi & \phi(\boldsymbol{x}_{p}) = \text{ } 4\pi\psi(\boldsymbol{x}_{p})  \\ \nonumber 
 & -  \int_{S} \left\{ \phi(\boldsymbol{x}) \left[ \frac{\partial{G(\boldsymbol{x},\boldsymbol{x}_{p})}}{\partial{n}} - \frac{\partial{G(\boldsymbol{x},\boldsymbol{x}_{0})}}{\partial{n}} \right] - \psi(\boldsymbol{x}) \left[ \frac{\partial{G_0(\boldsymbol{x},\boldsymbol{x}_{p})}}{\partial{n}} - \frac{\partial{G_0(\boldsymbol{x},\boldsymbol{x}_{0})}}{\partial{n}} \right] \right\} \text{ d}S(\boldsymbol{x})   \\ 
& + \int_{S} \left\{ \left[ G(\boldsymbol{x},\boldsymbol{x}_{p}) -G(\boldsymbol{x},\boldsymbol{x}_{0})\right] \frac{\partial{\phi(\boldsymbol{x})}}{\partial{n}} - \left[ G_0(\boldsymbol{x},\boldsymbol{x}_{p}) - G_0(\boldsymbol{x},\boldsymbol{x}_{0}) \right] \frac{\partial{\psi(\boldsymbol{x})}}{\partial{n}} \right\} \text{ d}S(\boldsymbol{x}).
\end{align}
This expression for the potential $\phi(\boldsymbol{x}_{p})$ at any point $\boldsymbol{x}_{p}$ in the solution domain given by Eq.~(\ref{eq:BRIEFfield}) contains no singular or near singular behavior and will give equally good precision irrespective of the location of the field point $\boldsymbol{x}_{p}$.

\nocite{*}
\bibliography{DH}

\end{document}